\documentclass[onecolumn,11pt,letterpaper]{usetex-onecol}

\usepackage{algorithm}
\usepackage{algpseudocode}
\usepackage{booktabs}
\usepackage{graphicx}
\usepackage{url}
\usepackage{color}
\usepackage{multirow}
\usepackage{amsmath}
\usepackage{amsthm}
\usepackage{xspace}
\usepackage{array}
\usepackage{enumitem}
\usepackage{epstopdf}
\usepackage[normalem]{ulem}
\usepackage{microtype}
\usepackage{subcaption}
\usepackage{multicol}
\usepackage{xcolor}
\DisableLigatures{encoding=*, family=*}

\usepackage{times}
\usepackage{breakurl}
\usepackage{color}
\usepackage{enumitem}
\usepackage{graphicx}
\usepackage{xspace}
\usepackage{epstopdf}
\usepackage{url}
\usepackage{cite}
\usepackage[colorlinks=true,linkcolor=black,citecolor=red,pdfborder={0 0 0}]{hyperref}
\usepackage{breakurl}
\usepackage{enumitem}
\usepackage{microtype}

\setlist{nolistsep}
\newcommand{\sysname}{{\sf ECPipe}\xspace}
\renewcommand{\paragraph}[1]{\smallskip\noindent {\bf #1}}

\begin{document}

\title{Repair Pipelining for Erasure-Coded Storage: Algorithms and Evaluation%
\thanks{An earlier version of this article appeared in \cite{li17}.  In this
extended version, we extend repair pipelining for hierarchical data centers and
multi-block repair operations.  We also implement and evaluate repair
pipelining in Hadoop~3.1.1 HDFS. 
This work was supported in part by the Research Grants Council of Hong Kong
(GRF 14216316 and AoE/P-404/18) and National Natural Science Foundation of
China (61872414, 61502191, and 61802365).  Corresponding author: Patrick P. C.
Lee (pclee@cse.cuhk.edu.hk)}}

\author{Xiaolu Li$^\dagger$, Zuoru Yang$^\dagger$, Jinhong Li$^\dagger$,
Runhui Li$^\dagger$, Patrick P. C. Lee$^\dagger$, Qun Huang$^\ddagger$,
Yuchong Hu$^\ast$\\
$^\dagger$The Chinese University of Hong Kong\\
$^\ddagger$Peking University\\
$^\ast$Huazhong University of Science and Technology}

\maketitle

\begin{abstract}
We propose {\em repair pipelining}, a technique that speeds up the repair
performance in general erasure-coded storage.  By carefully scheduling the
repair of failed data in small-size units across storage nodes in a pipelined
manner, repair pipelining reduces the single-block repair time to
approximately the same as the normal read time for a single block in
homogeneous environments.  We further design different extensions of repair
pipelining algorithms for heterogeneous environments and multi-block repair
operations.  We implement a repair pipelining prototype, called \sysname, and
integrate it as a middleware system into two versions of Hadoop Distributed
File System (HDFS) (namely HDFS-RAID and HDFS-3) as well as Quantcast File
System (QFS).  Experiments on a local testbed and Amazon EC2 show that repair
pipelining significantly improves the performance of degraded reads and
full-node recovery over existing repair techniques.
\end{abstract}

\section{Introduction}

Distributed storage systems rely on data redundancy to provide fault
tolerance, so as to maintain availability and durability.  
{\em Replication}, which is traditionally used by production
systems \cite{calder11,ghemawat03}, provides the simplest form of redundancy
by keeping identical copies of data in different storage nodes.  However, the
raw storage cost of replication is overwhelming, especially with the massive
scale of data we face today.  {\em Erasure coding} provides a low-cost
redundancy alternative that incurs significantly lower storage overhead than
replication at the same fault tolerance level \cite{weatherspoon02}.  
Today's distributed storage systems adopt erasure coding to protect data
against failures in clustered \cite{ford10,huang12,rashmi13} or
geo-distributed environments \cite{aguilera13,resch11,muralidhar14,chen17},
and reportedly save petabytes of storage \cite{huang12,muralidhar14}.
In a nutshell, erasure coding takes two configurable parameters $n$ and $k$
(where $k < n$) as input. It transforms $k$ fixed-size units (called 
{\em blocks}) of original data into a set of $n$ coded blocks of the same
size, such that any $k$ out of $n$ (coded) blocks can reconstruct all original
data; in other words, the original data remains available even if any $n-k$
blocks are failed (either lost or unavailable).  For example, if $n=14$ and
$k=10$ (the same parameters used in Facebook's erasure coding deployment
\cite{muralidhar14}), the storage overhead is only $1.4\times$, while
tolerating $n-k=4$ failed blocks.  In contrast, replication incurs $5\times$
storage overhead to tolerate the same number of failed blocks.  We elaborate
erasure coding in detail in \S\ref{subsec:basics}.  

Although achieving storage efficiency, erasure coding has a drawback of
incurring high repair penalty.  Specifically, in erasure-coded storage, the
repair of a single failed block needs to read
multiple available blocks for reconstruction; in other words, it reads more
available data than the size of a failed block.  This is in contrast to
replication, whose repair can be simply done by reading another replica that
is of the same size as the failed block.  The excessive data not only
increases the read time to failed data as opposed to normal reads, but also
consumes bandwidth resources that could otherwise be made available for other
foreground jobs \cite{rashmi13}.  Thus, erasure coding in practice is mainly
used for storing less frequently read (i.e., warm/cold) data that needs
long-term persistence \cite{huang12,andre14,muralidhar14}, while frequently
read (i.e., hot) data remains replicated for efficient access.  

To mitigate the repair penalty of erasure coding, prior studies propose new
constructions of erasure codes that significantly reduce the amount of repair
traffic (e.g.,
\cite{dimakis10,huang12,khan12,sathiamoorthy13,rashmi14,rashmi15,pamies16,vajha18});
in particular, the minimum-storage regenerating (MSR) codes
\cite{dimakis10,rashmi15,pamies16,vajha18} provably minimize the amount
of repair traffic subject to the minimum storage redundancy.  While the repair
time is effectively reduced due to the reduction of repair traffic, it remains
higher than the normal read time in general since the minimum size of repair
traffic remains larger than the size of the failed block.  In view of this, we
pose the following question: {\em Can we further reduce the repair time of
erasure coding to almost the same as the normal read time?} This creates
opportunities for applying erasure coding to hot data for high storage
efficiency, while preserving read performance.  

We present a general technique called {\em repair pipelining} to speed up the
repair performance in general erasure-coded storage.  
Its main idea is to decompose the repair of a block in small-size
units (called {\em slices}) and carefully schedule the repair of multiple
slices in a pipelined manner (analogous to wormhole routing \cite{ni93}),
so as to distribute the repair traffic and fully utilize the bandwidth
resources of storage nodes. 
Contrary to the conventional wisdom that the repair of erasure coding is a slow
operation, repair pipelining reduces the single-block repair time to almost
the same as the normal read time for a single available block, regardless of
coding parameters, in homogeneous environments where network links have
identical bandwidth limits.  Also, it provides different heuristics to
mitigate the single-block repair time in heterogeneous environments where
network links have different bandwidth limits.  Furthermore, it supports
various practical erasure codes that are adopted by today's production
systems, including the classical Reed-Solomon codes \cite{reed60} and the
recent Local Reconstruction Codes \cite{huang12}. 

We point out that the notion of repair pipelining has also been studied in
several publications by other researchers (e.g., \cite{huang15,xu19,bai19});
note that PUSH \cite{huang15} was published before the conference version
\cite{li17} of this paper.  PUSH \cite{huang15} addresses full-node recovery
by constructing a linear repair path through $k$ available blocks and
performing block-level repair pipelining.  For different block sizes (a.k.a.
request unit sizes \cite{huang15}), PUSH achieves $1/k$ of the repair time of
conventional repair in full-node recovery.  Thus, we do not claim that the
concept of repair pipelining is the main contribution of this paper.  Instead,
our contributions are to {\em demonstrate the viability of repair pipelining
in various deployment environments through in-depth analysis and prototype
evaluation}.  As we show in this paper, applying slice-level repair pipelining
is non-trivial, since its repair performance gain also depends on how the
slice-level repair sub-operations are scheduled.  We refer readers to
\S\ref{sec:related} for more detailed comparisons of our work with the related
approaches. 

To summarize, we make the following contributions. 
\begin{itemize}\itemsep=3pt
\item
We design repair pipelining to address two types of repair operations:
degraded reads and full-node recovery.  We show that repair pipelining reduces
the single-block repair time to almost the same as the normal read time for a
single available block in homogeneous environments.
\item
We extend repair pipelining to address heterogeneous environments and present
three extensions of repair pipelining algorithms: the first one allows parallel
reads of a repaired block when the bandwidth between the storage system and
the node that requests for the repaired block is limited; the second one finds
an optimal repair path for hierarchical data centers in which the cross-rack
link bandwidth is limited; the third one finds an optimal repair path across
storage nodes such that the repair time is minimized in a heterogeneous
environment where network links have arbitrary bandwidth limits. 
\item
We further extend repair pipelining for repairing multiple failed blocks
within the set of $n$ coded blocks.  We show that it reduces the multi-block
repair time to almost the same as the total normal read time for $f$ available
blocks in homogeneous environments, where $f$ is the number of failed blocks
being repaired. 
\item
We implement a repair pipelining prototype called \sysname, which runs as a
middleware system atop an existing storage system and performs repair
operations on behalf of the storage system.  
As a proof of concept, we integrate \sysname into two versions of Hadoop
Distributed File System (HDFS) \cite{shvachko10}, namely HDFS-RAID
\cite{fbHadoop} and Hadoop~3.1.1 HDFS (HDFS-3) \cite{hadoop3}, as well as
Quantcast File System (QFS) \cite{ovsiannikov13}. 
All the integrations only make minor changes (with no
more than 245 lines of code) to the code base of each storage system. 
The latest source code of our \sysname prototype is available at:
\href{http://adslab.cse.cuhk.edu.hk/software/ecpipe}%
{\bf http://adslab.cse.cuhk.edu.hk/software/ecpipe}.
\item
We evaluate repair pipelining on a local cluster and two geo-distributed
Amazon EC2 clusters (one in North America and one in Asia).  
We compare it with two existing repair approaches in which the single-block
repair time increases with $k$ (recall that $k$ is the number of blocks of
original data for encoding): 
(i) conventional repair that is used by classical Reed-Solomon codes
\cite{reed60} and achieves $O(k)$ single-block repair time, and (ii) the
recently proposed partial-parallel-repair (PPR) scheme \cite{mitra16}, which
achieves $O(\log_2 k)$ single-block repair time by parallelizing partial
repair operations in a hierarchical manner.  In contrast, repair pipelining
achieves $O(1)$ single-block repair time if there is a sufficiently large 
number of slices per block (i.e., independent of $k$).
Our experiments show that repair pipelining reduces the single-block repair
time by nearly 90\% and 70\% compared to conventional repair and PPR,
respectively. It also reduces the multi-block repair time by around 60\%
compared to conventional repair, as well as improves the repair performance in
HDFS-RAID, HDFS-3, and QFS deployments.  Furthermore, we show that our current
repair pipelining implementation in \sysname, by carefully parallelizing
slice-level repair sub-operations, achieves the highest performance for large
block sizes compared to several baseline repair pipelining implementations
(\S\ref{subsec:eval_rp}). 
\end{itemize}

The rest of the paper proceeds as follows.  
In \S\ref{sec:background}, we describe the basics of erasure coding and
motivate the repair problem.
In \S\ref{sec:design}, we present the design of repair pipelining. 
In \S\ref{sec:extensions}, we extend repair pipelining for heterogeneous
environments and multi-block repair operations. 
In \S\ref{sec:impl}, we present implementation details of \sysname and show
how it is integrated into existing open-source distributed storage systems. 
In \S\ref{sec:eva}, we present evaluation results. 
In \S\ref{sec:related}, we review related work, and finally, 
in \S\ref{sec:conclusions}, we conclude the paper. 

\section{Background and Motivation}
\label{sec:background}

We first present the basics of erasure coding and explain the repair problem.
We then motivate the need of minimizing the repair time in erasure-coded
storage.

\subsection{Basics}
\label{subsec:basics}

We consider a distributed storage system (e.g., GFS \cite{ghemawat03}, HDFS
\cite{shvachko10}, and Azure \cite{calder11}) that manages large-scale
datasets and stores files as fixed-size {\em blocks}, which form the basic
read/write units.  The block size is often large, ranging from 64\,MiB
\cite{ghemawat03} to 256\,MiB \cite{rashmi14}, to mitigate I/O seek overhead.
Erasure coding is applied to a collection of blocks.  Specifically, an erasure
code is typically configured with two integer parameters $(n,k)$, where 
$k < n$.  An $(n,k)$ code divides blocks into groups of $k$.  For every $k$
(uncoded) blocks, it encodes them to form $n$ coded blocks, such that any $k$
out of $n$ coded blocks can be decoded to the original $k$ uncoded blocks.
The set of $n$ coded blocks is called a {\em stripe}.  A large-scale storage
system stores data of multiple stripes, all of which are independently
encoded. The $n$ coded blocks of each stripe are distributed across $n$
distinct nodes to tolerate any $n-k$ node failures.  Most practical erasure
codes are {\em systematic}, such that $k$ of $n$ coded blocks are identical to
the original uncoded blocks and hence can be directly accessed without
decoding.  Nevertheless, our design treats both uncoded and coded blocks the
same, so we simply refer to them as ``blocks''.

Many erasure code constructions have been proposed in the literature (see
survey \cite{plank13} and \S\ref{sec:related}).  Among all erasure codes,
Reed-Solomon (RS) codes \cite{reed60} are the most popular erasure codes that
are widely deployed in production \cite{ford10,rashmi13,ovsiannikov13}.  There
are two key properties of RS codes: (i) {\em maximum distance separable
(MDS)}, meaning that RS codes can reconstruct the original $k$ uncoded blocks
from any $k$ out of $n$ coded blocks with the minimum storage redundancy
(i.e., $n/k$ times the original data size), and (ii) {\em general}, meaning
that RS codes support any $n$ and $k$ (provided that $k < n$).  

\begin{figure}[t]
\centering
\includegraphics[width=3in]{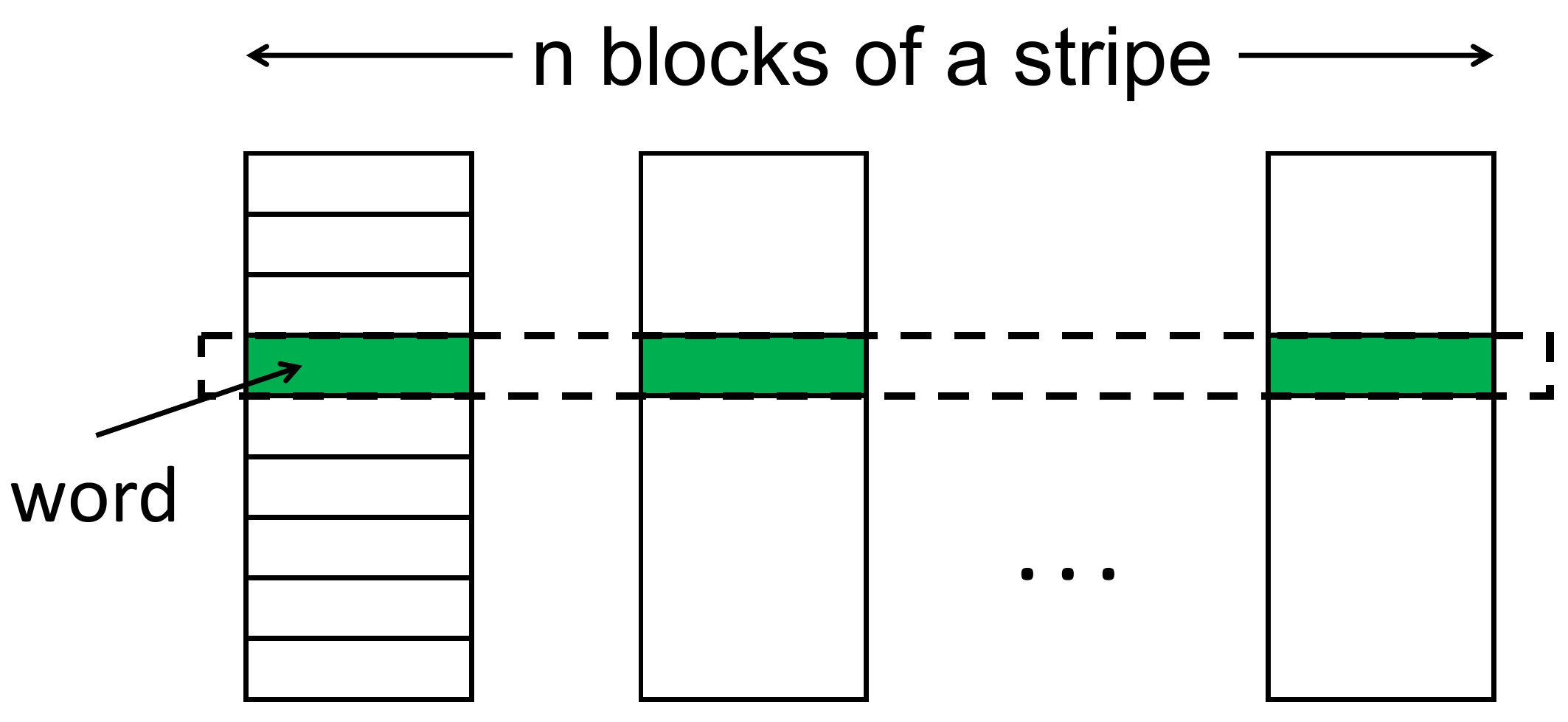}
\caption{In erasure coding, blocks are partitioned into words, such that words
at the same offset of each block of a stripe are encoded together.}
\label{fig:basicEC}
\end{figure}

Practical erasure codes (e.g., RS codes) often
satisfy {\em linearity}.  Specifically, for each stripe of an $(n,k)$ code,
let $\{B_1, B_2, \cdots, B_k\}$ denote any $k$ blocks of a stripe.  Any block
in the same stripe, say $B^*$, can be computed from a linear combination of
the $k$ blocks as $B^* = \sum_{i=1}^k a_i B_i$, where $a_i$'s ($1\le i\le k$)
are decoding coefficients specified by a given erasure code.  All additions
and multiplications are based on Galois Field arithmetic over $w$-bit units
called {\em words}; in particular, an addition is equivalent to bitwise XOR.
Note that the additions of $a_i B_i$'s are associative (i.e., the additions
can be in any order).  Some constraints may be applied; for example, RS codes
require $n\le2^w + 1$ \cite{plank09}.  Each block is partitioned into multiple
$w$-bit words, such that the words at the same offset of each block of a
stripe are encoded together, as shown in Figure~\ref{fig:basicEC}.

\subsection{Repair}
\label{subsec:repair}

In this paper, we focus on two types of repair operations in erasure-coded
storage: (i) {\em degraded reads} to temporarily unavailable blocks (e.g., due
to power outages, network disconnection, system maintenance, etc.) or lost
blocks that are yet recovered; and (ii) {\em full-node recovery} for restoring
all lost blocks of a failed node (e.g., due to disk crashes, sector errors,
etc.).  Each failed block (either lost or unavailable) is reconstructed in
a destination termed {\em requestor}, which can be a new node that replaces a
failed node, or a client that issues degraded reads.  Note that there may be
one or multiple requestors when multiple failed blocks are reconstructed.

Erasure coding triggers more repair traffic than the size of failed data to be
reconstructed.  For example, for $(n,k)$ RS codes, repairing a failed block
reads $k$ available blocks of the same stripe from other nodes (i.e., $k$
times the block size).  Some repair-friendly erasure codes (e.g.,
\cite{dimakis10,khan12,huang12,sathiamoorthy13,rashmi14,rashmi15,pamies16,vajha18})
are designed to reduce the repair traffic (see details in
\S\ref{sec:related}), but the amount of repair traffic
per block remains larger than the size of a block.  In distributed storage
systems, network bandwidth is often the most dominant factor in the repair
performance as extensively shown by previous work
\cite{dimakis10,silberstein14,mitra16} (see further justifications in
\S\ref{subsec:motivation}).  Thus, the amplification of repair traffic
implies the congestion at the downlink of the requestor, thereby increasing
the overall repair time.

\begin{figure}[t]
\centering
\begin{tabular}{c@{\hspace{0.5in}}c}
\includegraphics[width=2.3in]{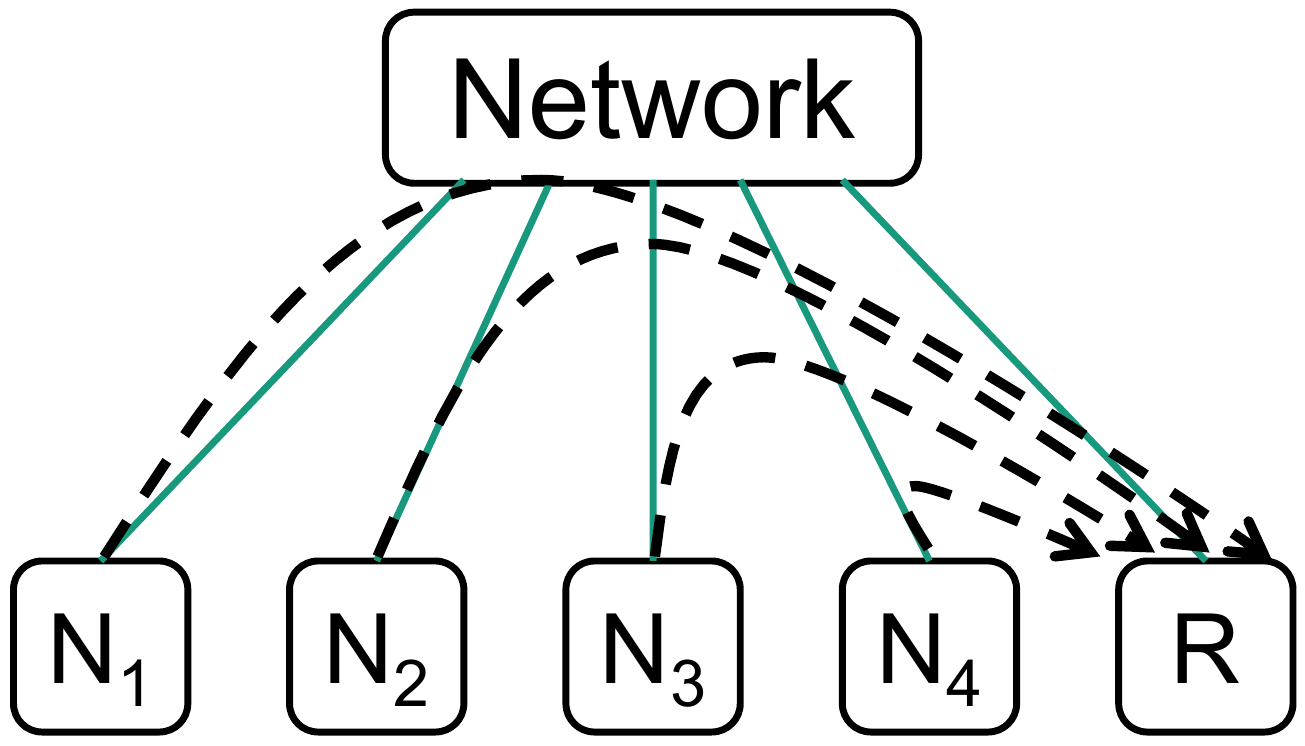}&
\includegraphics[width=2.3in]{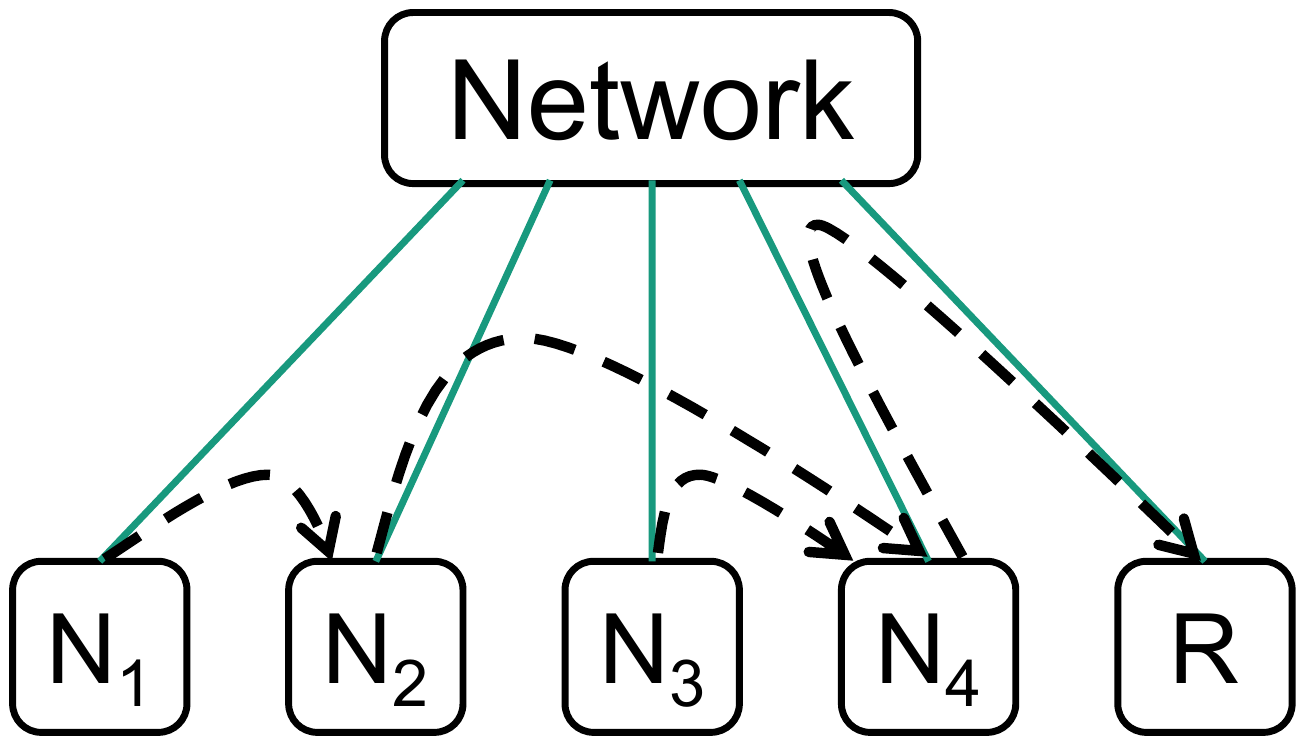}\\
\mbox{\small(a) Conventional repair}&
\mbox{\small(b) PPR}
\end{tabular}
\caption{Examples of conventional repair and PPR in a single-block repair.}
\label{fig:bgRec}
\end{figure}

To understand the repair penalty of erasure coding, we use RS codes as an
example and call this repair approach {\em conventional repair}.  Suppose that
a requestor $R$ wants to repair a failed block $B^*$.  It can be done by
reading $k$ available blocks from any $k$ working nodes, called
{\em helpers}. Without loss of generality, let $R$ contact $k$ helper nodes
$N_1$, $N_2$, $\cdots$, $N_k$, which store available blocks $B_1$, $B_2$,
$\cdots$, $B_k$, respectively.  To make our
discussion clear, we divide the repair process into {\em timeslots}, such that
only one block can be transmitted across a network link in each timeslot.
Figure~\ref{fig:bgRec}(a) shows how conventional repair works for $k = 4$.
Since $R$ needs to retrieve the $k$ blocks $B_1$, $B_2$, $\cdots$, $B_k$, all
$k$ transmissions must traverse the downlink of $R$.  Overall, the repair in
Figure~\ref{fig:bgRec}(a) takes {\em four} timeslots.  In general,
conventional repair needs $k$ timeslots to repair a failed block. 

Conventional repair can address the repair of multiple concurrently
failed blocks in the same stripe.  Suppose that there are $f \le n-k$ failed
blocks in a stripe (i.e., fault tolerance is still maintained).  Our goal is
to repair the $f$ failed blocks in $f$ requestors, each of which stores a
reconstructed block.  The multi-block repair can be done by dedicating one of
the $f$ requestors to retrieve $k$ available blocks from $k$ helper nodes.
Since the dedicated requestor has sufficient information to reconstruct all
original uncoded data, it can also reconstruct all $f$ failed blocks.  Thus,
it can locally store one of the reconstructed blocks and send the $f-1$
reconstructed blocks to the other $f-1$ requestors.  The number of timeslots
for a multi-block repair is $k+f-1$ timeslots. 

A drawback of conventional repair is that the bandwidth usage distribution is
highly skewed: the downlink of the requestor is highly congested, while the
links among helpers are not fully utilized.  PPR \cite{mitra16} builds on the
linearity and addition associativity of erasure coding by decomposing a repair
operation into multiple partial operations that are distributed across all
helpers.  This distributes bandwidth usage across the links of helpers.
Figure~\ref{fig:bgRec}(b) shows how PPR repairs $B^*$ for $k = 4$.
In the first timeslot, $N_2$ and $N_4$ receive blocks $a_1 B_1$ and $a_3 B_3$
from $N_1$ and $N_3$, respectively.  Since the transmissions use different
links, they can be done simultaneously in a single timeslot.
In the second timeslot, $N_2$ combines the received $a_1 B_1$ and its locally
stored block $B_2$ to obtain $a_1 B_1 + a_2 B_2$ and sends it to $N_4$.
In the third timeslot, $N_4$ combines all received blocks and its own
block $B_4$ to obtain $a_1 B_1 + a_2 B_2 + a_3 B_3 + a_4 B_4$, and sends it to
$R$.  This hierarchical approach reduces the overall single-block repair time
to only {\em three} timeslots.  In general, PPR needs $\lceil\log_2(k +
1)\rceil$ timeslots to repair a failed block.  Note that how to generalize PPR
for repairing multiple failed blocks in a stripe is still an unexplored issue.

\subsection{Motivation}
\label{subsec:motivation}

Although PPR reduces the single-block repair time, the bandwidth usage
distribution remains not fully balanced; for example, the downlink of $N_4$ in
Figure~\ref{fig:bgRec}(b) still carries more repair traffic than other links.
Thus, the repair time is still bottlenecked by the link with the most repair
traffic.  This motivates us to design a new repair scheme that can more
efficiently utilize bandwidth resources, with the primary goal of minimizing
the repair time.

Minimizing the repair time is critical to both availability and durability. In
terms of availability, field studies show that transient failures (i.e., no
data loss) account for over 90\% of failure events \cite{ford10}.  Thus, most
repairs are expected to be degraded reads rather than full-node recovery.
Since degraded reads are issued when clients request unavailable
data, achieving fast degraded reads not only improves availability but is also
critical for meeting customer service-level agreements \cite{huang12}.  In
terms of durability, minimizing the repair time also minimizes the window of
vulnerability.  By recovering any failed block in a timely manner, we maintain
the redundancy level for fault-tolerant storage.  This avoids any
unrecoverable data loss if the number of failed blocks exceeds the tolerable
limit (i.e., $n-k$ blocks for an $(n,k)$ code).  

Our work targets distributed storage environments in which network bandwidth
is the bottleneck.  Although modern data centers now scale to high network
speeds, they are typically shared by a mix of application workloads.  Thus,
the network bandwidth available for repair tasks is often throttled
\cite{huang12,silberstein14}.  Also, modern data centers often have
hierarchical network topologies by organizing nodes in racks, in which the
cross-rack link bandwidth is limited (e.g., due to replica writes
\cite{chowdhury13} or compute job traffic \cite{ahmad14,jalaparti15}). 
To tolerate rack failures, data centers distribute each stripe 
across racks \cite{ford10,huang12,sathiamoorthy13,rashmi14}.  Thus, the repair
of any failed block inevitably retrieves available blocks from other racks and
triggers cross-rack transmissions.  The repair performance will be
bottlenecked by the limited cross-rack link bandwidth.

\section{Repair Pipelining}
\label{sec:design}

We present the design of repair pipelining.  We first state our goals and
assumptions (\S\ref{subsec:goals}).  We then explain how repair pipelining
addresses degraded reads (\S\ref{subsec:degraded}) and full-node recovery
(\S\ref{subsec:recovery}). 

\subsection{Goals and Assumptions}
\label{subsec:goals}

Repair pipelining also exploits the linearity and addition associativity
of erasure codes as in PPR \cite{mitra16}, yet it parallelizes the repair
across helpers in an inherently different way.  It focuses on (i) eliminating
bottlenecked links (i.e., no link transmits more traffic than others) and (ii)
effectively utilizing bandwidth resources during a repair (i.e., links should
not be idle for most times), so as to ultimately minimize the single-block
repair time to the normal read time for a single block in homogeneous
environments where all links have the same bandwidth.  

Repair pipelining is mainly designed for speeding up the repair of a single
failed block per stripe, which is much more common than the repair of multiple
failed blocks per stripe in practice \cite{huang12,rashmi13} (e.g., over 98\%
of repair cases are single-block repair operations).
Optimizing a single-block repair is also the main design goal of most existing 
repair-friendly erasure codes that aim to minimize the amount of repair traffic
\cite{dimakis10,khan12,huang12,sathiamoorthy13,rashmi14,rashmi15,pamies16,vajha18}. 
In this section, we focus on studying the single-block repair for one stripe
and multiple stripes.  The former occurs when a requestor issues a degraded
read to an unavailable block; the latter occurs when all lost data of a single
failed node is recovered in one or multiple requestors in full-node recovery.

If a stripe has multiple failed blocks, we can also extend repair pipelining
to trigger a multi-block repair, which we show incurs less repair time than
conventional repair (\S\ref{subsec:repair}).  See \S\ref{subsec:multifail} for
details. 

Repair pipelining does not design new repair-friendly erasure codes that
minimize the repair traffic (the same assumption is made in PPR
\cite{mitra16}).  Instead, each repair of a single failed block still reads
$k$ available blocks as in conventional repair, yet it spreads the repair
traffic across all $k$ helpers so as to fully utilize bandwidth resources.
The failed block can then be reconstructed as a linear combination of the $k$
available blocks.  

Some repair-friendly codes, including locally repairable codes
\cite{huang12,sathiamoorthy13} and Rotated RS codes \cite{khan12}, work by
reconstructing a failed block through a linear combination of fewer than $k$
available blocks.  In this case, we can also combine repair pipelining with
such repair-friendly codes to reduce the single-block repair time while
preserving their repair traffic savings.  We evaluate such a combination in
\S\ref{subsec:eval_local}.  An interesting open question is to augment repair
pipelining with the optimal repair of general repair-friendly codes (e.g.,
regenerating codes \cite{dimakis10}), so as to simultaneously reduce the
single-block repair time and minimize the repair traffic.  We pose this
question as a future work. 

\subsection{Degraded Reads}
\label{subsec:degraded}

We first study how repair pipelining reconstructs a single block of a stripe
in a requestor in a degraded read.  We start with a na\"ive approach.
Specifically, we arrange $k$ helpers and the requestor as a linear path,
i.e., $N_1 \rightarrow N_2 \rightarrow \cdots \rightarrow N_k \rightarrow R$.
At a high level, to repair a failed block $B^*$, $N_1$ sends $a_1 B_1$ to $N_2$.
Then $N_2$ combines $a_1 B_1$ with its own block $B_2$ and sends $a_1 B_1 +
a_2 B_2$ to $N_3$.  The process repeats, and finally, $N_k$ sends $R$ the
combined result, which is $B^*$.  The whole repair incurs $k$ transmissions
that span across $k$ different links. Thus, there is no bottlenecked link. 
However, this na\"ive approach underutilizes bandwidth resources, since there
is only one block-level transmission in each timeslot.  The whole repair still
takes $k$ timeslots, same as in conventional repair (\S\ref{subsec:repair}). 

\begin{figure}[t]
\centering
\includegraphics[width=4in]{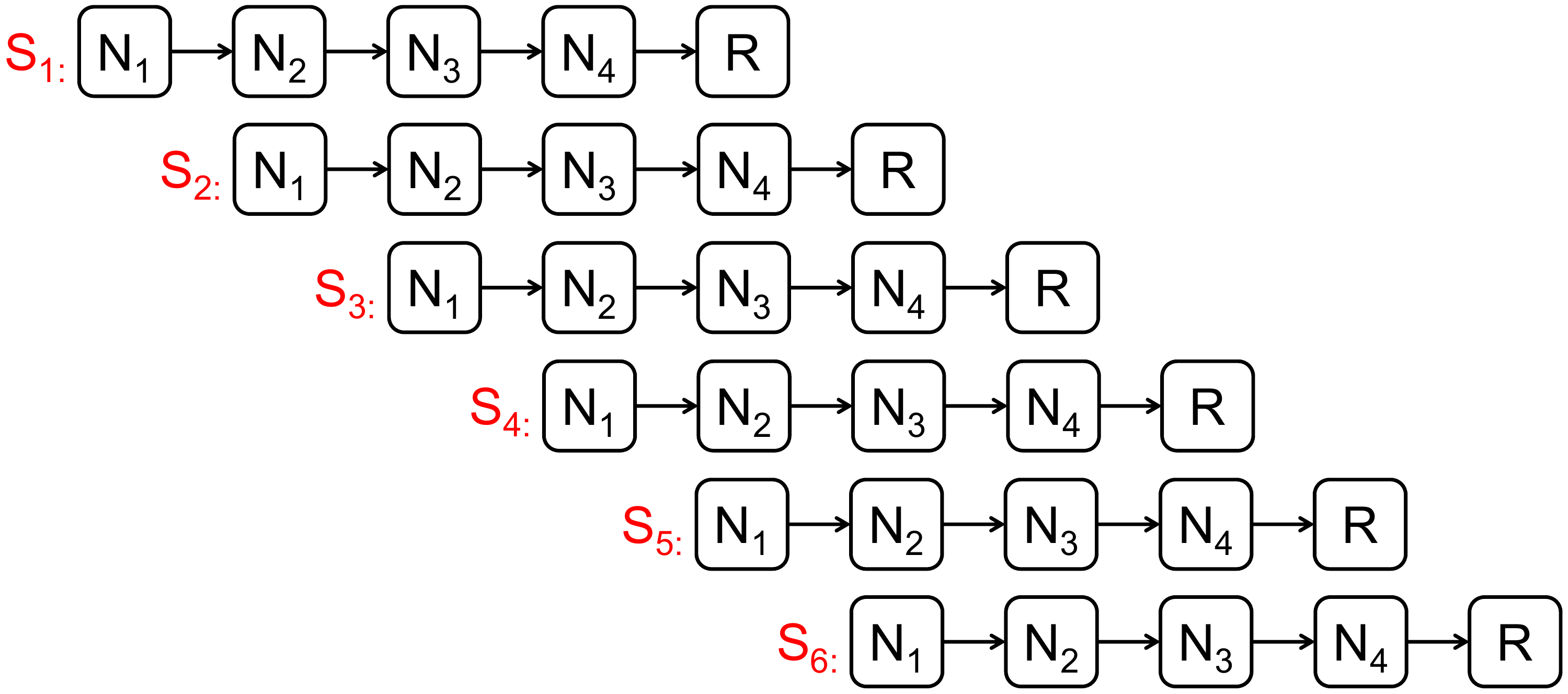}
\caption{Repair pipelining with $k = 4$ and $s = 6$.}
\label{fig:designPip}
\end{figure}

Thus, repair pipelining decomposes the repair of a block into the repair of a
set of $s$ small fixed-size units called {\em slices} $S_1, S_2, \cdots, S_s$.
It also partitions each block $B_i$ ($1\le i\le k$) into $s$ slices $B_{i,1},
B_{i,2}, \cdots, B_{i,s}$.  It pipelines the repair of each slice through the
linear path.  To repair the first slice $S_1$, $N_1$ sends $a_1 B_{1,1}$ to
$N_2$, $N_2$ sends $a_1 B_{1,1} + a_2 B_{2,1}$ to $N_3$, and so on.  Note that
when $N_2$ sends the slice $a_1 B_{1,1} + a_2 B_{2,1}$ to $N_3$, $N_1$ can
start the repair of the second slice $S_2$ by sending $a_1 B_{1,2}$ to $N_2$
without interfering in the transmission from $N_2$ to $N_3$.  Thus, we can
parallelize the slice-level transmissions.  Each slice-level transmission over
a link only takes $\frac{1}{s}$ timeslots.  Figure~\ref{fig:designPip} shows
how repair pipelining works for $k=4$ and $s=6$. 

A slice can have an arbitrarily small size, provided that Galois Field
arithmetic can be performed (\S\ref{subsec:basics}).  For RS codes, the
minimum size of a slice is a $w$-bit word; if $w=8$, a word denotes a byte. 
On the other hand, practical distributed storage systems store data in
large-size blocks, typically 64\,MiB or even larger (\S\ref{subsec:basics}).
Since a coding unit (i.e., word) has a much smaller size than a read/write
unit (i.e., block), we can parallelize a block-level repair operation into
more fine-grained slice-level repair sub-operations.  Having smaller-size
slices improves parallelism, yet it increases the overhead of issuing many
requests for transmitting slices over the network.  We study the impact of the
slice size in \S\ref{sec:eva}. 

We analyze the time complexity of repair pipelining.  Here, we neglect the
overheads due to computation and disk I/O, which we assume cost less time than
network transmission; in fact, both computation and disk I/O operations can
also be executed in parallel with network transmission in actual
implementation (\S\ref{sec:impl}).  Each slice-level transmission over a link
takes $\frac{1}{s}$ timeslots.  The repair of each slice takes $\frac{k}{s}$
timeslots to traverse the linear path, and $N_1$ starts to transmit the last
slice after $\frac{s-1}{s}$ timeslots.  Thus, the whole repair time, which is
given by the total number of timeslots to transmit all slices through the
linear path, is $\frac{s-1+k}{s} = 1+\frac{k-1}{s}$ timeslots.  In practice,
$k$ is of moderate size to avoid large coding overhead \cite{plank09} (e.g.,
$k=12$ in Azure \cite{huang12} and $k=10$ in Facebook \cite{rashmi13}), while
$s$ can be much larger (e.g., $s=$~2,048 for 32\,KiB slices in a 64\,MiB
block).  Thus, we have $1+\frac{k-1}{s} \to 1$, as $s$ is sufficiently large. 

Repair pipelining connects multiple helpers as a linear path, so its repair
performance is bottlenecked by the presence of poorly performed links/helpers
(i.e., stragglers).  We emphasize that any repair scheme of erasure coding
faces the similar problem, as it retrieves available data from multiple helpers
for data reconstruction; for example, conventional repair for $(n,k)$ MDS
codes needs to retrieve the available data from $k$ helpers.  We address the
straggler problem by taking into account heterogeneity and bypassing
stragglers via helper selection (\S\ref{subsec:weighted}). Also, if any helper
fails during an ongoing repair, the progress of repair pipelining will be
stalled.  In this case, we restart the whole repair process with a new set of
available helpers and trigger a multi-block repair (\S\ref{subsec:multifail}).

\subsection{Full-Node Recovery}
\label{subsec:recovery}

We now study how repair pipelining addresses a multi-stripe repair (one failed
block per stripe) when recovering a full-node failure.  As the stripes are
independently encoded, we can parallelize the multiple single-stripe repair
operations.  However, since each repair involves a number of helpers, if one
helper is chosen in many repair operations of different stripes, it will
become overloaded and slow down the overall repair performance.  In practice,
each stripe is stored in a different set of storage nodes spanning across the
network.  Our goal is to distribute the load of a multi-stripe repair across
all helpers as evenly as possible.  

We adopt a simple greedy scheduling approach for the selection of helpers.
For each node in the storage system, repair pipelining keeps track of a
timestamp indicating when the node was last selected as a helper for a
single-stripe repair.  To repair a failed block of a stripe, we select $k$ out
of $n-1$ available helpers in the stripe that have the smallest timestamps; in
other words, the $k$ selected helpers are the least recently selected ones in
previous requests.  Choosing $k$ out of the $n-1$ helpers can be done in
$O(n)$ time using the quick select algorithm \cite{hoare61} (based on repeated
partitioning of quick sort).  We use a centralized coordinator to manage the
selection process (\S\ref{sec:impl}).  Our greedy scheduling emphasizes
simplicity in deployment.  We can also adopt a more sophisticated approach by
weighting node preferences in real time \cite{mitra16}. 

Unlike the degraded read scenario, the multiple reconstructed blocks can be
stored in multiple requestors.  Under this condition, the gain of repair
pipelining over conventional repair decreases, as the latter can also
parallelize the repair across multiple requestors.  Nevertheless, our
evaluation indicates that repair pipelining
still provides repair performance improvements (\S\ref{sec:eva}).  

Note that the number of requestors that can be selected and the choices of
requestors may depend on various deployment factors \cite{mitra16}. In
this work, we assume that the requestors are selected offline in advance. 

\section{Extensions}
\label{sec:extensions}

We now extend the basic design of repair pipelining in \S\ref{sec:design} to
address three different heterogeneous settings, in which the links of a
distributed storage system no longer have identical bandwidth: (i) a requestor
can read slices from multiple helpers in parallel in which the link bandwidth
from the storage system to the requestor is limited (\S\ref{subsec:parallel});
(ii) we arrange the linear path of $k$ helpers in a hierarchical data center
with limited cross-rack link bandwidth (\S\ref{subsec:rackaware}); and (iii)
we solve a weighted path selection problem to find an optimal path of $k$
helpers that maximizes the repair performance in a heterogeneous environment
where network links have arbitrary bandwidth (\S\ref{subsec:weighted}).
Finally, we extend repair pipelining to address a multi-block repair
(\S\ref{subsec:multifail}).

\subsection{Parallel Reads}
\label{subsec:parallel}

In the basic design of repair pipelining, a requestor always reads slices
from one helper.  This may lead to last-mile congestion.  For example, a
client (requestor) sits at the network edge and accesses a cloud storage
system that is far from the client.  We propose a {\em cyclic version} of
repair pipelining that allows a requestor to read slices from multiple helpers.

\begin{figure}[t]
\centering
\includegraphics[width=4.3in]{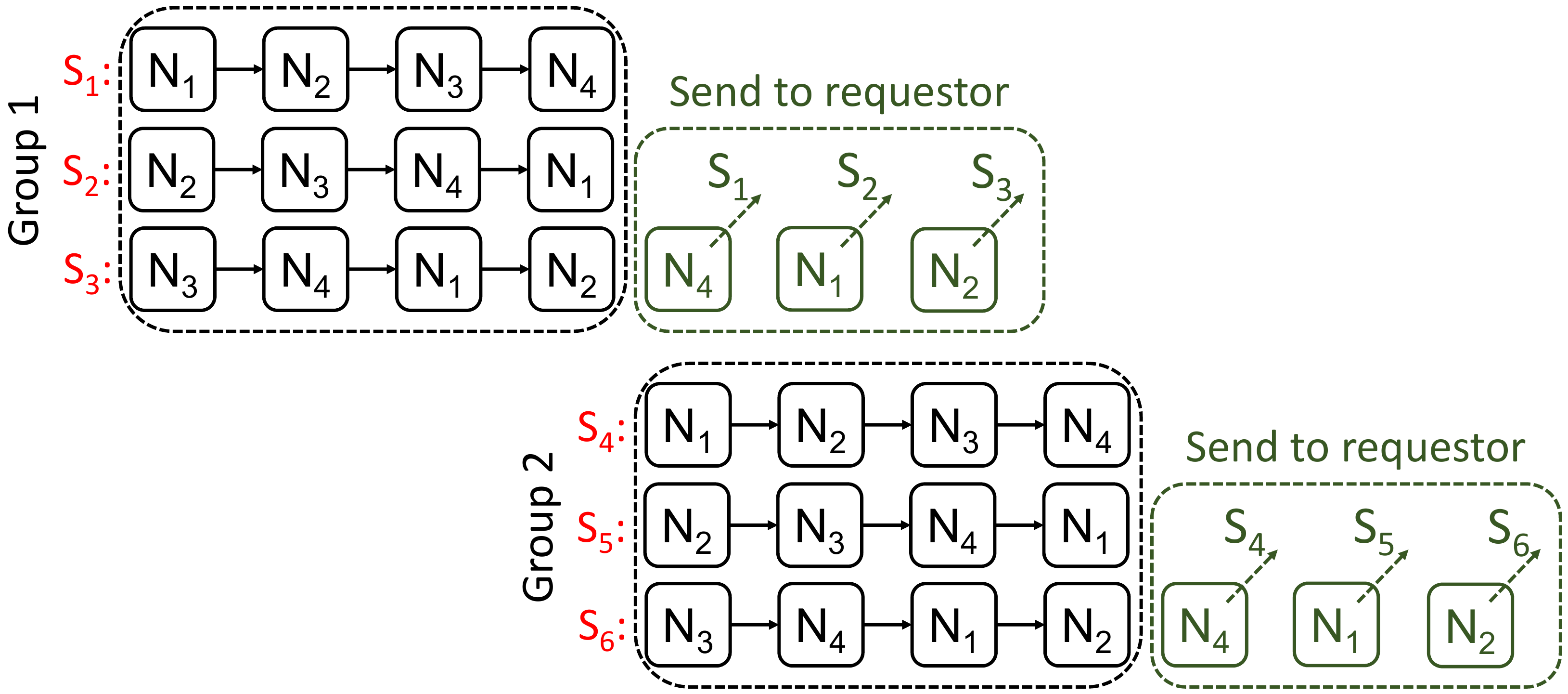}
\caption{Cyclic version of repair pipelining with $k = 4$ and $s = 6$.}
\label{fig:designCyc}
\end{figure}

We now describe the cyclic version.  Our discussion assumes that all links are
homogeneous, and transmitting a block size of data over a link takes one
timeslot.  The cyclic version again divides a failed block into $s$ fixed-size
slices $S_1$, $S_2$, $\cdots$, $S_s$, and repairs each slice through some
linear path to eliminate any bottlenecked link. However, it now maps the $k$
helpers $N_1$, $N_2$, $\cdots$, $N_k$ into different {\em cyclic paths} that
can be cycled from $N_k$ through $N_1$.  Specifically, it partitions the $s$
slices into $\lceil\tfrac{s}{k-1}\rceil$ groups, each of which has $k-1$ slices
(the last group has fewer than $k-1$ slices if $s$ is not divisible by $k-1$).
The repair of each group of slices is then performed in two phases.  Without
loss of generality, we only consider how to repair the first group $S_1$,
$S_2$, $\cdots$, $S_{k-1}$.  In the first phase, repairing each slice $S_i$
($1\le i\le k-1$) traverses through the cyclic path $N_i \rightarrow N_{i+1}
\rightarrow \cdots N_k \rightarrow N_1 \rightarrow \cdots N_{i-1}$.  We repair
all slices through different cyclic paths simultaneously, and each slice-level
transmission takes $\tfrac{1}{s}$ timeslots.  The first phase can be done in
$\tfrac{k-1}{s}$ timeslots.  In the second phase, the last helper of each
cyclic path delivers the repaired slice to the requestor.  The second phase is
also done in $\tfrac{k-1}{s}$ timeslots.  Figure~\ref{fig:designCyc} shows the
cyclic version for $k=4$ and $s=6$.

Note that we can start repairing the slices of the next group simultaneously
while we deliver the repaired slices for the current group.  Specifically,
while $k-1$ helpers simultaneously transmit slices for the repair in the next
group, there is one idle helper that can transmit the repaired slice for the
current group to the requestor.   They can be done together in $\tfrac{k-1}{s}$
timeslots.

We analyze the time complexity of the cyclic version under the homogeneous
link assumption.  We only consider the case where $s$ is divisible by $k-1$,
while the same result can be derived otherwise.  Repairing each group of
slices takes $\tfrac{2(k-1)}{s}$ timeslots, and the repair of the last group
starts after $(\tfrac{s}{k-1}-1)\tfrac{k-1}{s}$ timeslots.  The whole repair
time is $(\tfrac{s}{k-1}-1)\tfrac{k-1}{s} + \tfrac{2(k-1)}{s} = 1 + \tfrac{k-1}{s}
\rightarrow 1$, as $s$ is sufficiently large.

Note that the cyclic version now allows a requestor to read slices from $k-1$
helpers.  If the repair bottleneck lies in the network transfer from the
helpers to the requestor, our evaluation shows that the cyclic version
significantly outperforms the basic design of repair pipelining
(\S\ref{sec:eva}).

\subsection{Hierarchy Awareness}
\label{subsec:rackaware}

We extend repair pipelining to address hierarchical network topologies. Here, we
focus on rack-based data centers, which organize storage nodes in racks, such
that the available cross-rack link bandwidth is much more limited than the
available inner-rack link bandwidth (\S\ref{subsec:motivation}).  {\em Our
goal is to not only minimize the single-block repair time, but also minimize
the amount of cross-rack repair traffic incurred for the single-block repair}.
Note that our analysis is also applicable for geo-distributed data centers
\cite{aguilera13,resch11,muralidhar14,chen17}, where storage nodes span
different geographical regions and the cross-region bandwidth is much more
limited than the inner-region bandwidth (\S\ref{subsec:eval_ec2}). 

\paragraph{Background:}
Recent studies (e.g., \cite{hu17,prakash18,hou19}) have designed optimal
rack-aware erasure codes that provably minimize the amount of cross-rack
repair traffic in a single-block repair from an information theoretical
perspective.  Some studies (e.g., CAR \cite{shen16} and LAR \cite{xu19}) focus
on RS codes and propose cross-rack-aware repair strategies that minimize the
amount of cross-rack repair traffic under RS codes.  In all such designs, the
idea is to place multiple blocks of a stripe per rack, such that a
single-block repair first computes a partially repaired block (which is a
linear combination of the available blocks within a rack), followed by
aggregating the partially repaired blocks across racks; note that each rack is
required to store at most $n-k$ blocks, so as to provide a single-rack fault
tolerance for an $(n,k)$ code.  

As a rack failure now makes multiple blocks unavailable, such a hierarchical
block placement trades rack-level fault tolerance for the reduction of
cross-rack repair traffic.  We can measure the reliability trade-off based on
the commonly used mean-time-to-data-loss (MTTDL) measure via Markov analysis.
The MTTDL measure depends on both failure rates (for both independent and
correlated node failures) and repair rates.  It is shown by Hu {\em et al.}
\cite{hu17} that hierarchical block placement can achieve a higher MTTDL than
flat block placement through minimizing the cross-rack repair traffic,
provided that (i) the rack-based data center has limited cross-rack link
bandwidth or (ii) correlated node failures (or rack failures) are less
frequent than independent node failures; note that condition~(ii) is also
justified in practical geo-distributed data centers \cite{muralidhar14}.  We
refer readers to the study \cite{hu17} for the detailed reliability analysis.
Instead of designing new erasure codes, we extend repair pipelining with rack
awareness for general erasure codes under the hierarchical block placement.
Since repair pipelining reduces the single-block repair time, we expect that
the storage reliability (in MTTDL) further improves.

\begin{figure}[t]
\centering
\begin{tabular}{c@{\hspace{0.5in}}c}
\includegraphics[width=2in]{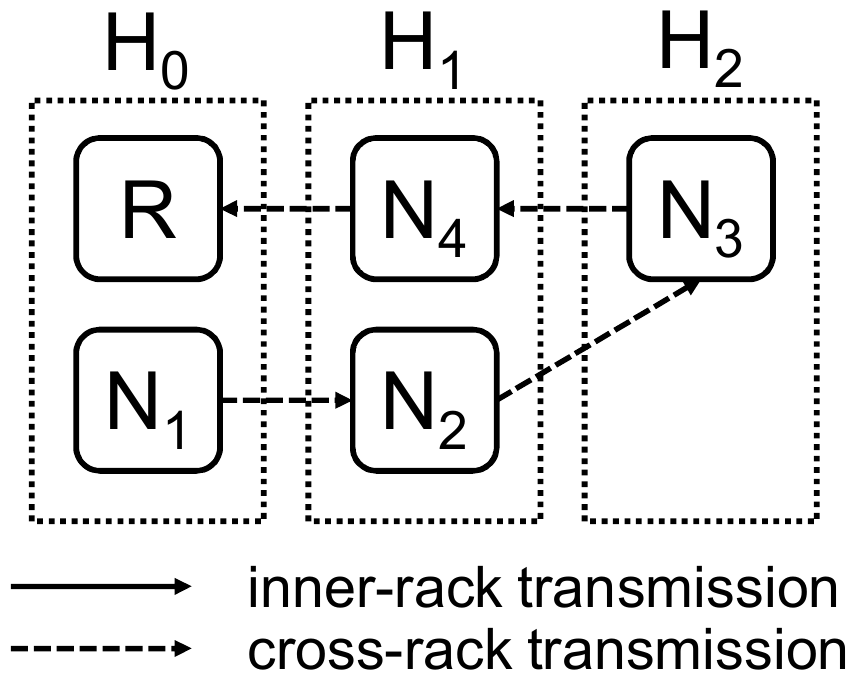}&
\includegraphics[width=2in]{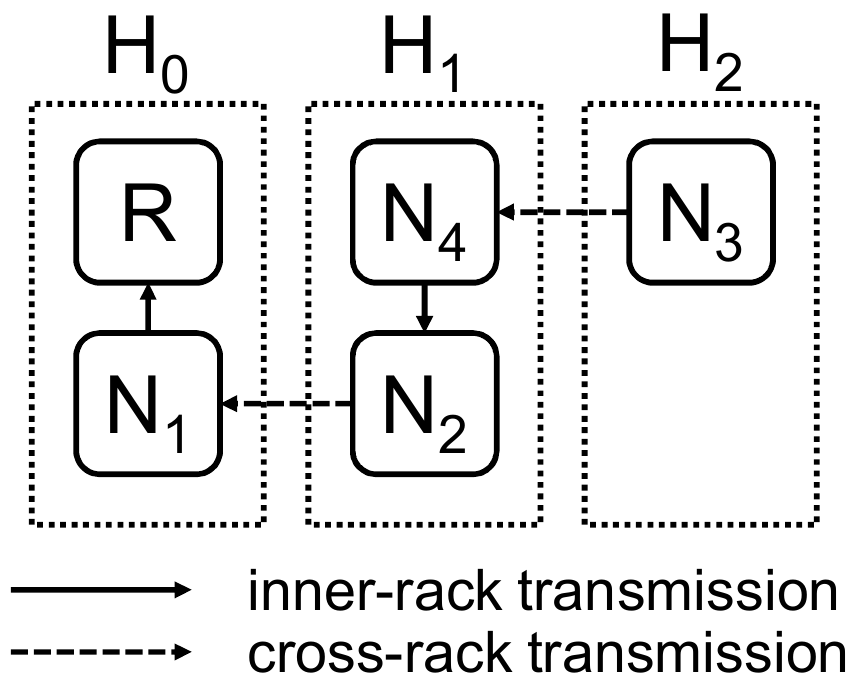}\\
\mbox{\small(a) No rack awareness}&
\mbox{\small(b) With rack awareness}
\end{tabular}
\caption{Repair pipelining with rack-aware path selection.} 
\label{fig:rackaware}
\end{figure}

\paragraph{Algorithm:}
Our idea is that the linear path of $k$ helpers in repair pipelining should
limit cross-rack transmissions.  Figure~\ref{fig:rackaware}(a) shows a linear
path of $k=4$ helpers that are randomly ordered without rack awareness.  In
this example, the middle rack has two simultaneous incoming cross-rack
transmissions (i.e., $N_1 \rightarrow N_2$ and $N_3 \rightarrow N_4$), thereby
creating congestion at the downlink bandwidth of the middle rack. 

\begin{algorithm}[!t]
\caption{Rack-Aware Path Selection}
\label{alg:rackaware}
\small
\begin{algorithmic}[1]
\item[{\bf Input:} data center topology]
\item[{\bf Output:} path $P$]
\State identify the racks $\{H_i\}$ where the requestor $R$ and $n-1$
available helpers reside
\State let $H_0$ be the rack containing $R$ (and some helpers)
\State let $H_1, H_2, \cdots, H_h$ be the remote racks containing the remaining
helpers, sorted by the number of helpers in a rack in descending order
\State $P = R$
\State $i = 0$
\While{$P$ has fewer than $k$ helpers}
  \For{each helper $N$ in $H_i$}
    \State $P = N \rightarrow P$
    \If{$P$ has $k$ helpers}
       \State break the while loop 
    \EndIf 
  \EndFor
  \State $i = i + 1$
\EndWhile
\State {\bf return} $P$
\end{algorithmic}
\end{algorithm}

To make repair pipelining rack-aware, we require that the linear path of $k$
helpers has at most one incoming transmission and at most one outgoing
transmission for each rack, while minimizing the total number of cross-rack
transmissions.  Algorithm~\ref{alg:rackaware} shows the pseudo-code of the
rack-aware path selection.  Specifically, to repair a failed block of a
stripe, we identify a requestor $R$ and the remaining $n-1$ available helpers
of the stripe.  Let $H_i$ ($i\ge 0$) denote a rack where either $R$ or any
helper resides, such that $H_0$ denotes the rack that contains $R$ (and
possibly other helpers), $H_1, H_2, \cdots, H_h$ denote a total of $h$ 
{\em remote racks} that do not contain $R$ but contain the remaining helpers.
Without loss of generality, we sort the number of helpers in the remote racks
$H_i$'s ($1\le i\le h$) in descending order, where $|H_1| \ge |H_2| \ge \cdots
\ge |H_h|$, and $|H_i|$ ($1\le i\le h$) denotes the number of helpers in
$H_i$.  We first initialize the linear path $P$ with only the requestor $R$
(Line~4).  We then iteratively append a helper in $H_i$ to $P$, starting from
$i=0$, until $P$ has $k$ helpers for reconstructing the failed block
(Lines~5-14).  Figure~\ref{fig:rackaware}(b) shows a linear path for $k=4$
with rack-aware path selection. 

Our rationale is that we prefer to append all helpers that are co-located with
$R$ in $H_0$ to $P$, so as to involve only inner-rack transmissions.  Also,
when choosing helpers from the remote racks $H_1, H_2, \cdots, H_h$, we
prefer to append as many helpers as possible in one rack to $P$, so as to
minimize the number of remote racks to be accessed.  Thus, we first choose
helpers from $H_1$, followed by $H_2$, and so on.  By minimizing the
number of remote racks being accessed for a single-block repair, we also
minimize the amount of cross-rack repair traffic under RS codes (see CAR
\cite{shen16}).  Based on the analysis in \S\ref{sec:design}, the single-block
repair time still approaches one timeslot, while a timeslot here refers to the
time of transmitting one block over a cross-rack link. 

\paragraph{Remarks:}  A recent work LAR \cite{xu19} solves for a minimum
spanning tree that takes the network distances among nodes as input and
minimizes the cross-rack repair traffic in the network core of a hierarchical
topology.  For the special case where all nodes share the identical network
distance to the network core, LAR can also return a linear path with the
minimum cross-rack repair traffic as Algorithm~\ref{alg:rackaware}.  Unlike
LAR, which uses network distances as input, Algorithm~\ref{alg:rackaware} uses
only the block locations in different racks (as in CAR \cite{shen16}) as input
to find the linear path.  Repair pipelining takes one step further to reduce
the single-block repair time based on the returned linear path. 

\subsection{Weighted Path Selection}
\label{subsec:weighted}

We now study a more diverse heterogeneous setting in which the link bandwidth
can have any arbitrary value.  In the following, we extend repair pipelining to
solve a {\em weighted path selection} problem.  Here, we focus on degraded
reads, and discuss how we address full-node recovery. 

\paragraph{Formulation:}
Recall that for a single-block repair, repair pipelining transmits a number of
slices along a linear path of $k$ helpers, say $N_1\rightarrow N_2 \rightarrow
\cdots \rightarrow N_k\rightarrow R$.  Suppose that the link bandwidth is
different across links.  If the number of slices is sufficiently large, then
the slices are transmitted in parallel through the path
(Figure~\ref{fig:designPip}), and the performance of repair pipelining will be
bottlenecked by the link with the minimum available bandwidth along the path.
To minimize the single-block repair time, we should find a path that maximizes
the minimum link bandwidth.

To repair a failed block of a stripe, we need to find $k$ out of $n-1$
available helpers of the same stripe as the failed block, and also find the
sequence of link transmissions so that the path along the $k$ selected helpers
and the requestor minimizes the single-block repair time.  Specifically, there
are a total of $n$ nodes, including the $n-1$ available helpers and the
requestor.  We associate a {\em weight} with each (directed) link from one
node to another node, such that a higher weight implies a longer transmission
time along the link.  For example, the weight can be represented by the
inverse of the link bandwidth obtained by periodic measurements on link
utilizations \cite{chowdhury13}.   Then our objective is to find a path of
$k+1$ nodes (i.e., $k$ selected helpers and the requestor) that minimizes the
maximum link weight of the path.  Here, we focus on link weights, and the same
idea is applicable if we associate weights with nodes.  Any straggler is
assumed to be associated with a large weight, so it will be excluded from the
selected path.

To solve the above problem, a na\"ive approach is to perform a
brute-force search on all possible candidate paths.  However, there are a
total of $\tfrac{(n - 1)!}{(n - 1 - k)!}$ permutations, and the brute-force
search becomes computationally expensive even for moderate sizes of $n$ and
$k$.  Since the link weights vary over time, the path selection should be done
quickly on-the-fly based on the measured link weights.

\begin{algorithm}[!t]
\caption{Weighted Path Selection}
\label{alg:hetero}
\small
\begin{algorithmic}[1]
\item[{\bf Input:} link weights]
\item[{\bf Output:} optimal path $P^*$]
\Procedure {Main}{}
\State $P = R$
\State $P^* = $ {\bf null}
\State $w^* = \infty$
\State $\mathcal{N} = $ set of $n-1$ available helpers
\State {\sc ExtendPath}
\State {\bf return} $P^*$
\EndProcedure
\Function {ExtendPath}{}
  \If {$P$.length $< k + 1$}
	\For {each node $N \in \mathcal{N}$ not in $P$}
	  \If {weight($N$, beginning node of path $P$) $< w^*$}
		\State $P = N \rightarrow P$
		\State {\sc ExtendPath}
		\State remove $N$ from $P$
	  \EndIf
	\EndFor
  \Else
	\State $P^* = P$
	\State $w^* = $ maximum link weight of $P$
  \EndIf
\EndFunction
\end{algorithmic}
\end{algorithm}

\paragraph{Algorithm:} We present a fast yet optimal algorithm that quickly
identifies an optimal path.  The algorithm builds on brute-force search to
ensure that all candidate paths are covered, but eliminates the search of
infeasible paths.  Our insight is that if a link $L$ has a weight larger than
the maximum weight of an optimal path candidate that is currently found, then
we no longer need to search for the paths containing link $L$, since the
maximum weight of any path containing $L$ must be larger than the maximum
weight of the optimal path candidate.

Algorithm~\ref{alg:hetero} shows the pseudo-code of the weighted path
selection algorithm.  Let $P$ be the path that we currently consider, $P^*$ be
the optimal path candidate that we have found, $w^*$ be the maximum link
weight of $P^*$, and $\mathcal{N}$ be the set of $n-1$ available helpers.  We
first initialize a path $P$ with only the requestor $R$ (Line~2), such that
$R$ will be the end node of $P$. We also initialize $P^*$, $w^*$, and
$\mathcal{N}$ (Lines~3-5).  We call the recursive function {\sc ExtendPath}
(Line~6) and finally return the optimal path $P^*$ (Line~7).

The function {\sc ExtendPath} recursively extends $P$ by one helper in
$\mathcal{N}$ and appends the helper to $P$ if the link weight
from the node to the current beginning node of $P$ is less than $w^*$;
otherwise, the path containing the link cannot minimize the maximum link
weight as argued above.  Specifically, the algorithm appends $N\in\mathcal{N}$
to $P$ if the current path length is less than $k+1$ and the weight from $N$
to the beginning node of $P$ is less than $w^*$ (Lines~10-13).  It calls 
{\sc ExtendPath} again to consider candidate paths that now include
$N\rightarrow P$ (Line~14).  It then removes $N$ from $P$ (Line~15), and tries
other nodes in $\mathcal{N}$.  If the length of $P$ is now $k+1$, it implies
that all of its links have weight less than $w^*$, so we update $P$ as the new
optimal path $P^*$ and $w^*$ as the maximum link weight of $P^*$
(Lines~19-20).

Algorithm~\ref{alg:hetero} significantly reduces the search time.
We evaluate the search time for (14,10) codes using Monte-Carlo
simulations over 1,000 runs on a machine with 3.7\,GHz Intel Xeon E5-1620 v2 CPU
and 16\,GiB memory.  The brute-force search takes 27\,s on average, while
Algorithm~\ref{alg:hetero} reduces the search time to only 0.9\,ms.

\paragraph{Remarks:} To address full-node recovery (\S\ref{subsec:recovery}),
we apply Algorithm~\ref{alg:hetero} to each stripe.  If we apply greedy
scheduling on helper selection, we simply substitute $\mathcal{N}$ with the
set of $k$ selected helpers.  Note that the brute-force search for the optimal
path on the $k$ selected helpers remains expensive, since it considers $k!$
permutations on the sequence of link transmissions along the path. Thus,
Algorithm~\ref{alg:hetero} still significantly saves the search time in this
case.

We emphasize that Algorithm~\ref{alg:hetero} should {\em not} be
viewed as a generalization of our rack-aware path selection in
Algorithm~\ref{alg:rackaware} (\S\ref{subsec:rackaware}) as both algorithms
target different problem settings:  Algorithm~\ref{alg:rackaware} specifically 
minimizes the number of cross-rack transmissions, while
Algorithm~\ref{alg:hetero} minimizes the maximum link weight of the linear
path of helpers.  Nevertheless, we can still apply Algorithm~\ref{alg:hetero}
in a geo-distributed environment (\S\ref{subsec:eval_ec2}). 

\subsection{Multi-Block Repair}
\label{subsec:multifail}

Finally, we show how repair pipelining simultaneously reconstructs multiple
failed blocks of a stripe and reduces the multi-block repair time.  Here, we
only focus on homogeneous environments, and discuss how we can address the
heterogeneous environments. 

We first define the notation.  Let $f$, where $1\le f\le n-k$, be the number
of failed blocks of a stripe for an $(n,k)$ code, and $B_1^*, B_2^*, \cdots,
B_f^*$ be the failed blocks to be reconstructed.  Let $R_1, R_2, \cdots, R_f$
be the $f$ requestors where the failed blocks are reconstructed.  Before
issuing the repair, we first identify $k$ helpers of the stripe (say, 
$N_1, N_2, \cdots, N_k$) and their $k$ corresponding available blocks (say,
$B_1, B_2, \cdots, B_k$, respectively).  Each failed block $B_j^*$ 
($1\le j\le f$) can be reconstructed via the linear combination $B_j^* =
\sum_{i=1}^k a_{i,j} B_i$, where $a_{i,j}$'s $(1\le i\le k$, $1\le j\le f$)
are the decoding coefficients specified by a given erasure code.  

A straightforward multi-block repair approach is to invoke repair pipelining
for a single-block repair over a linear path of $k$ helpers as described in
\S\ref{subsec:degraded} $f$ times, one for each failed block.  Thus, the
multi-block repair time approaches $f$ timeslots under the homogeneous link
assumption, where a timeslot is the time for transmitting one block over a
network link.  However, each helper needs to read its locally stored block 
for each single-block repair, so it reads $f$ times its locally stored block
in total.  In the following, we re-design a multi-block repair approach in
which each helper needs to read its locally stored block only once. 

As in \S\ref{subsec:degraded}, we start with a na\"ive pipelining approach
that realizes a multi-block repair {\em without slicing}, and show its
limitations.  Specifically, we arrange the $k$ helpers in a linear path, i.e.,
$N_1\rightarrow N_2\rightarrow \cdots\rightarrow N_k$, and connect $N_k$ to
all $f$ requestors $R_1, R_2, \cdots, R_f$.  To repair the $f$ failed blocks
$\{B_1^*, B_2^*, \cdots, B_f^*\}$,  $N_1$ uses its own block $B_1$ to compute
a set of $f$ blocks $\{a_{1,1}B_1, a_{1,2}B_1, \cdots, a_{1,f}B_1\}$, where
each $a_{1,j}B_1$ ($1\le j\le f$) is an input term to the linear combination
for reconstructing $B_j^*$.  $N_1$ sends the set of $f$ blocks to $N_2$.  Then
$N_2$ combines the received blocks with its own block $B_2$ and sends a new
set of $f$ blocks $\{a_{1,1}B_1 + a_{2,1}B_2, a_{1,2}B_1 + a_{2,2}B_2, \cdots,
a_{1,f}B_1 + a_{2,f}B_2\}$ to $N_3$.  The process repeats, and finally $N_k$
reconstructs all $f$ failed blocks $\{B_1^*, B_2^*, \cdots, B_f^*\}$ and sends
them to the $f$ requestors.  Note that each of the $k$ helpers reads its
own block only once.  For the total repair time, recall that a block-level
transmission over a network link takes one timeslot. Thus, the whole repair
incurs $f\times k$ timeslots, including $f(k-1)$ timeslots from $N_1$ to $N_k$
and $f$ timeslots from $N_k$ to all $f$ requestors.  From this example, we
observe that this na\"ive pipelining approach is even worse than conventional
repair (which takes $k+f-1$ timeslots as shown in \S\ref{subsec:repair}).

\begin{figure}[t]
\centering
\includegraphics[width=3.9in]{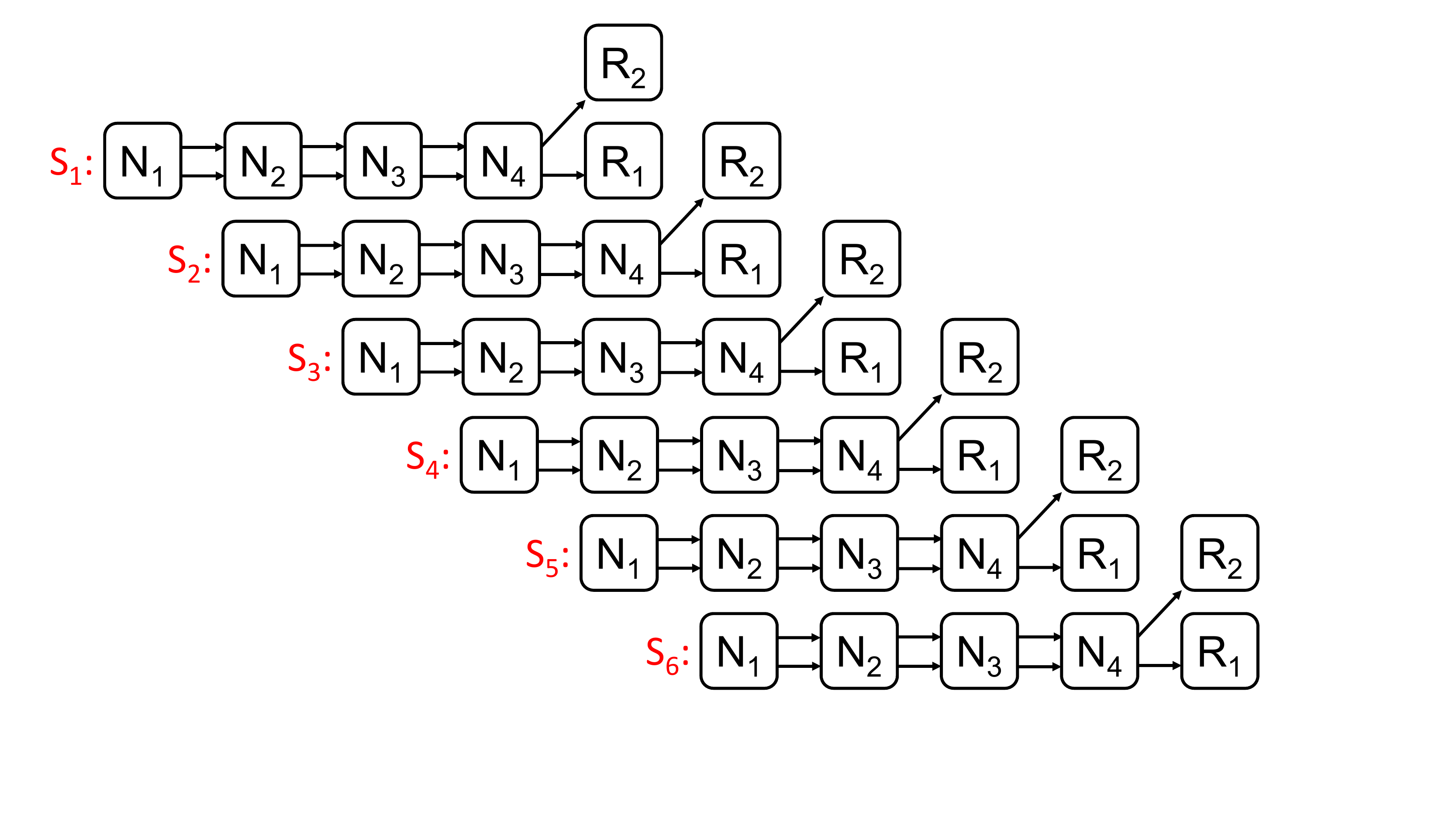}
\caption{Repair pipelining for a multi-block repair with $k=4$, $s=6$, and
$f=2$.}
\label{fig:designMulPip}
\end{figure}

We now extend the above na\"ive pipelining approach with slicing and show how
repair pipelining works for a multi-block repair.  Repair pipelining
decomposes each failed block $B_j^*$ ($1\le j\le f$) into $s$ slices denoted
by $S_{j,1}, S_{j,2}, \cdots, S_{j,s}$.  It pipelines the repair of the first
set of $f$ slices of the $f$ failed blocks (i.e., $S_{1,1}, S_{2,1}, \cdots,
S_{f,1}$) through a linear path, followed by the second set of $f$ slices
(i.e., $S_{1,2}, S_{2,2}, \cdots, S_{f,2}$), and so on.  In general, each
helper pipelines the repair of $f$ slices at the same offset of the $f$ failed
blocks along a linear path.  Each set of $f$ slices will be reconstructed at
$N_k$ (i.e., the last helper of the linear path), which then forwards the
reconstructed slices to the $f$ requestors.  Figure~\ref{fig:designMulPip}
shows how repair pipelining works for $k=4$, $s=6$, and $f=2$.  Again, each of
the $k$ helpers reads its locally stored block only once during the repair.  

We now analyze the time complexity of repair pipelining for repairing $f$
failed blocks under the homogeneous link assumption.  Again, we assume that
the overheads due to computation and disk I/O are negligible compared to
network transmission (\S\ref{subsec:degraded}).  To repair a set of $f$ slices
along a linear path, each helper sends $f$ slices to the next helper, or to
all $f$ requestors for the last helper $N_k$.  Thus, each transmission now
takes $\tfrac{f}{s}$ timeslots.  Following the analysis in
\S\ref{subsec:degraded}, the total repair time of repairing $f$ failed blocks
is $(s-1+k)\times\tfrac{f}{s} = f(1 + \tfrac{k-1}{s})$ timeslots, which
approaches $f$ timeslots if $s$ is sufficiently large.  Thus, repair
pipelining always incurs less repair time than conventional repair (which
takes $k+f-1$ timeslots).  

\paragraph{Remarks:}  For a heterogeneous environment where network links have
arbitrary bandwidth (\S\ref{subsec:weighted}), we discuss two possible
solutions to realize a multi-block repair.  One solution is to extend our
proposed design, in which we aggregate all $f$ requestors as one big
requestor, and assign a weight from each of the $n-f$ available helpers to the
big requestor.  Then we find an optimal linear path that minimizes the maximum
link weight as in \S\ref{subsec:weighted}.  An alternative solution is to call
a single-block repair for each of the $f$ failed blocks and find an optimal
path for each single-block repair as in \S\ref{subsec:weighted}.  We pose the
analysis for the possible solutions as future work. 

\section{Implementation}
\label{sec:impl}

We have implemented a prototype called \sysname to realize repair pipelining.
\sysname runs as a middleware atop an existing distributed storage system and
performs repair operations on behalf of the storage system.  Moving the repair
logic to \sysname greatly reduces changes to the code base of the storage
system in order to realize new repair techniques; in the meantime, we can
focus on optimizing \sysname to maximize the repair performance gain.  We have
integrated \sysname with three open-source distributed storage systems, namely
HDFS-RAID \cite{fbHadoop}, HDFS-3 \cite{hadoop3}, and QFS
\cite{ovsiannikov13}.  Both HDFS-RAID and HDFS-3 are written in Java, while
QFS is written in C++.  Our \sysname prototype is mostly written in C++, and
the parts for the integration into HDFS-RAID and HDFS-3 are in Java. Our
\sysname prototype has around 6,000 lines of code.  The latest source code of
\sysname is available at: 
\href{http://adslab.cse.cuhk.edu.hk/software/ecpipe}%
{\bf http://adslab.cse.cuhk.edu.hk/software/ecpipe}.

\subsection{Background of HDFS-RAID, HDFS-3, and QFS}

We first provide the background details of HDFS-RAID, HDFS-3, and QFS. In
particular, we describe how they support erasure coding. 

\paragraph{HDFS-RAID}:  HDFS-RAID is an erasure coding extension of
Hadoop~0.20 HDFS.  In this work, we choose Facebook's HDFS-RAID implementation
\cite{fbHadoop}.  Specifically, the original HDFS comprises a {\em NameNode}
for storage management and multiple {\em DataNodes} for actual storage.
HDFS-RAID deploys a {\em RaidNode} atop HDFS for erasure coding management.
It performs {\em offline encoding} (i.e., asynchronously in the background),
in which HDFS initially stores data in DataNodes as fixed-size blocks (64\,MiB
by default) with replication, and the RaidNode later encodes replicated blocks
into coded blocks via MapReduce \cite{dean04}.  

The RaidNode also checks for any failed (lost or corrupted) coded block by
verifying block checksums.  It repairs any failed block being detected, either
by itself in local mode or via a MapReduce job in distributed mode.  Both
modes will issue reads to $k$ available blocks of the same stripe in parallel
from HDFS, reconstruct the failed block, and write back to HDFS.  HDFS-RAID
also provides a RAID file system client to access coded blocks.  For a
degraded read to a failed block, the RAID file system reads $k$ available
blocks of the same stripe in parallel and reconstructs the failed block.

\paragraph{HDFS-3:} HDFS-3 (Hadoop~3.1.1 HDFS) \cite{hadoop3} includes erasure
coding in HDFS storage by design.  Unlike HDFS-RAID, HDFS-3 performs 
{\em online encoding} (i.e., on the write path), in which an HDFS client
performs encoding before writing data to storage.  Specifically, the HDFS
client first writes data into $k$ data buffers (with the default size of
1\,MiB) and encodes them into $n-k$ parity buffers.  It then appends the $n$
buffers into $n$ blocks in different DataNodes.  Compared to
offline encoding in HDFS-RAID, online encoding in HDFS-3 removes the extra I/O
costs of reading and encoding the currently stored data blocks. However, it
now moves the encoding overhead to the client, which performs encoding and
writes the data and parity blocks to HDFS storage.

The NameNode monitors any failed blocks via the periodic block reports issued
from DataNodes.  If a failed block exists, the NameNode assigns a repair task
to a DataNode, which issues parallel reads to $k$ available blocks from other
DataNodes, reconstructs the failed block, and writes the reconstructed block
back to HDFS-3. 

\paragraph{QFS:}  QFS stores all data in erasure-coded format and currently
supports (9,6) RS codes \cite{reed60}. Similar to HDFS-3, QFS performs online
encoding.  Specifically, a QFS client writes data into six 1\,MiB buffers.
It then encodes the six 1\,MiB buffers into three 1\,MiB parity buffers, and
appends the nine 1\,MiB buffers to nine data and parity blocks (the default
block size is 64\,MiB) that are stored in different storage nodes (called
{\em ChunkServers}).  To repair any failed block, a ChunkServer retrieves six
available blocks from other ChunkServers for reconstruction.

\subsection{\sysname Design}

\begin{figure}[t]
\centering
\includegraphics[width=4in]{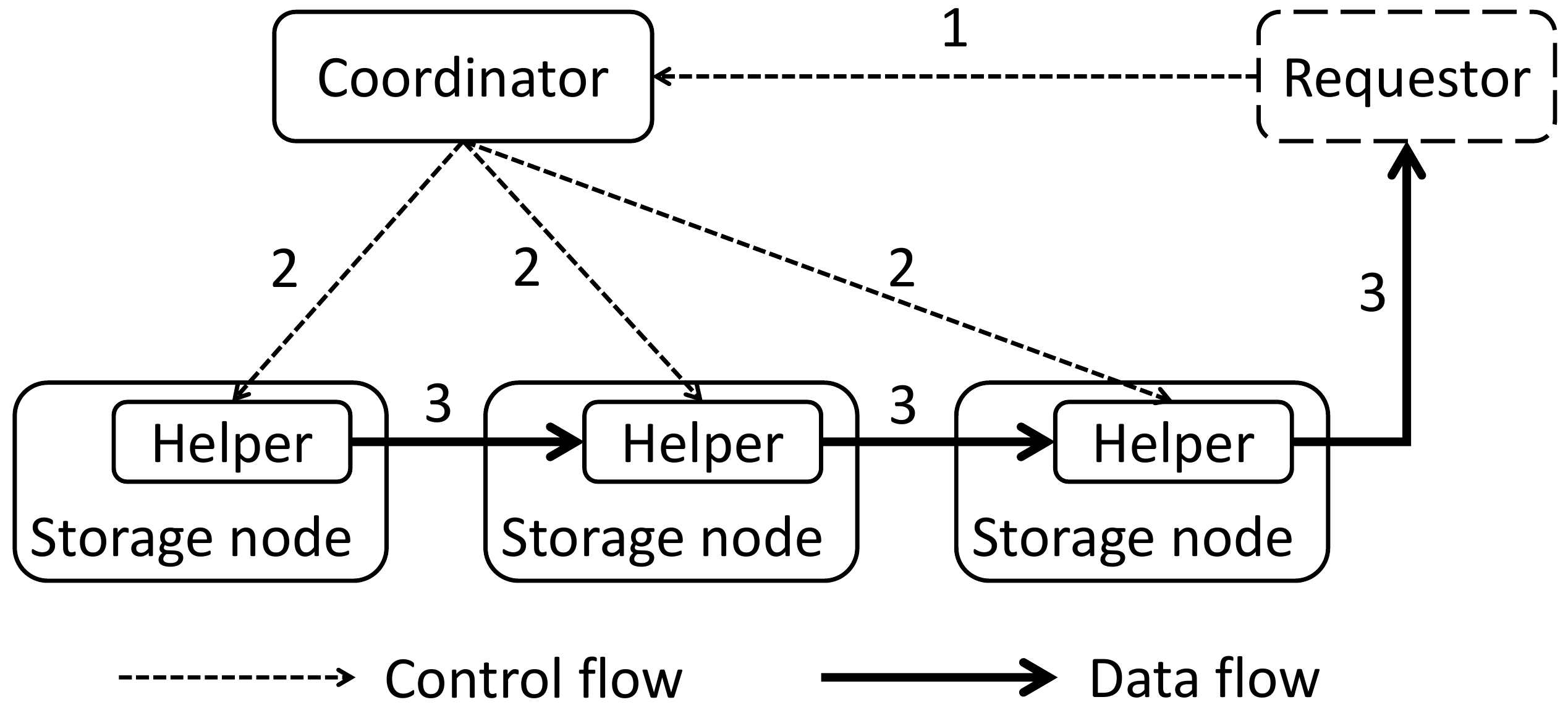}
\caption{\sysname architecture.}
\label{fig:impl}
\end{figure}

Figure~\ref{fig:impl} shows the \sysname architecture.  It uses a 
{\em coordinator} to manage the repair operation between a requestor and
multiple helpers.  \sysname runs atop a distributed storage system.  To repair
a failed block, the storage system creates a requestor instance, which sends a
repair request with the failed block ID to the coordinator (step~1).  The
coordinator uses the failed block ID to identify the locations of $k$
available blocks of the same stripe. It notifies all helpers with the block
locations (step~2). The helpers retrieve the blocks, perform repair pipelining
in slices, and deliver the repaired slices to the requestor (step~3).  

Note that if there are multiple failed blocks in a stripe, the storage system
creates multiple requestor instances, each of which issues the failed block ID
of one of the failed blocks to the coordinator.  Again, the coordinator
selects $k$ helpers to perform a multi-block repair via repair pipelining, so
that the failed blocks are reconstructed in multiple requestors. 

We integrate \sysname with a storage system in three aspects.  First, we
implement the requestor as a class (in C++ and Java) that can be instantiated
by the storage system to reconstruct failed blocks.  For HDFS-RAID, the
requestor is created in either the RaidNode or the RAID file system client;
for HDFS-3 and QFS, it is created by the storage node that starts a repair
operation. Second, we implement each helper as a daemon that is co-located
with each storage node to directly read the locally stored blocks.  Our
insight is that HDFS-RAID, HDFS-3, and QFS all store a block in the
underlying native file system as a plain file and use the block ID to form
the file name.  Thus, each helper can directly read the stored blocks via the
native file system.  This eliminates the need of helpers to fetch data through
the distributed storage system routine.  It not only reduces the burden of
metadata management of the distributed storage system, but also improves the
repair performance (\S\ref{subsec:eval_dfs}).  Finally, the coordinator needs to
access both the block locations and the mappings of each block to its stripe.
For HDFS-RAID, we retrieve the information from the RaidNode; for HDFS-3, we
retrieve the information from the NameNode; for QFS, we retrieve the
information from a storage node when it starts a repair operation. 

To simplify our implementation, \sysname uses Redis \cite{redis} to pipeline
slices across helpers.  Each helper maintains an in-memory key-value store
based on Redis, and uses the client interface of Redis to transmit slices
among helpers.  In addition, each helper performs disk I/O, network transfer,
and computation via multiple threads for performance speedup.  Adding \sysname
into HDFS-RAID, HDFS-3, and QFS only requires changes of around 110, 245, and
180 lines of code, respectively.

To provide fair comparisons (\S\ref{sec:eva}), we also implement conventional
repair (\S\ref{subsec:repair}) and PPR \cite{mitra16} under the same \sysname
framework, by only changing the transmission flow of data during a repair.

\section{Evaluation}
\label{sec:eva}

We conduct experiments on both a local cluster and Amazon EC2.  We show that
repair pipelining outperforms both conventional repair and PPR \cite{mitra16}
under various settings.  
We further show that our repair pipelining implementation in \sysname
outperforms different baseline repair pipelining implementations.

\subsection{Evaluation on a Local Cluster}
\label{subsec:eval_local}

\begin{figure*}[t]
\begin{tabular}{@{\ }c@{\ }c@{\ }c}
\includegraphics[width=2.2in]{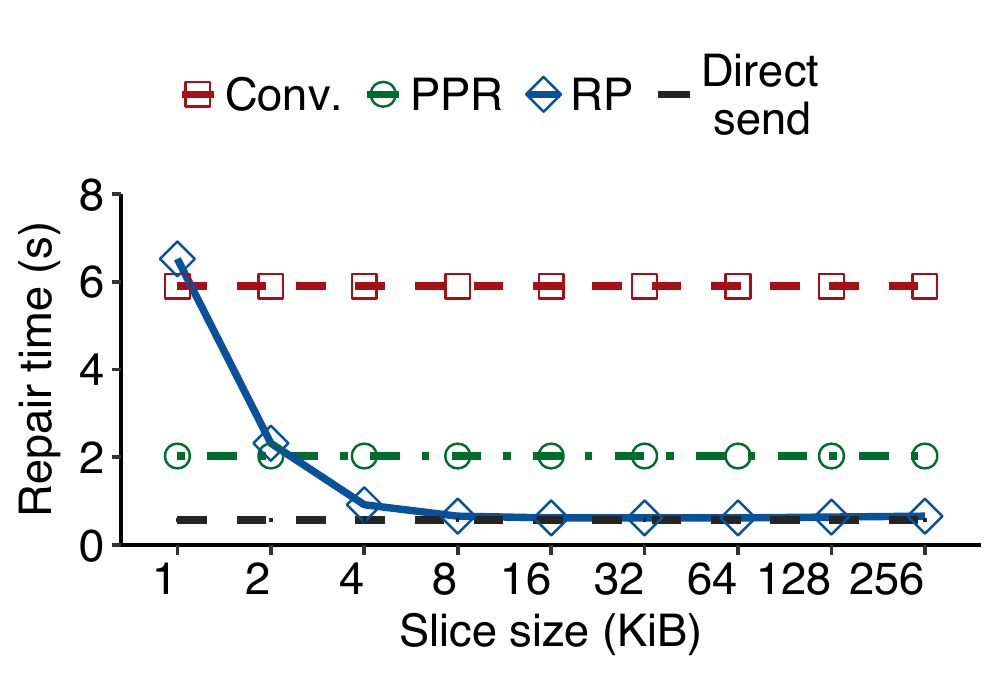} &
\includegraphics[width=2.2in]{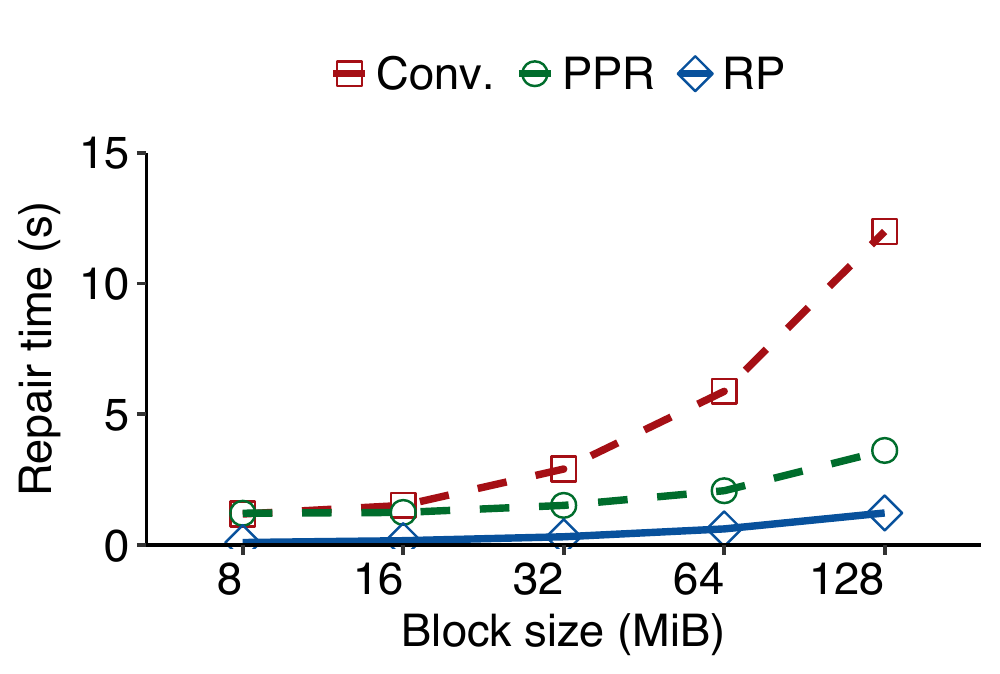} &
\includegraphics[width=2.2in]{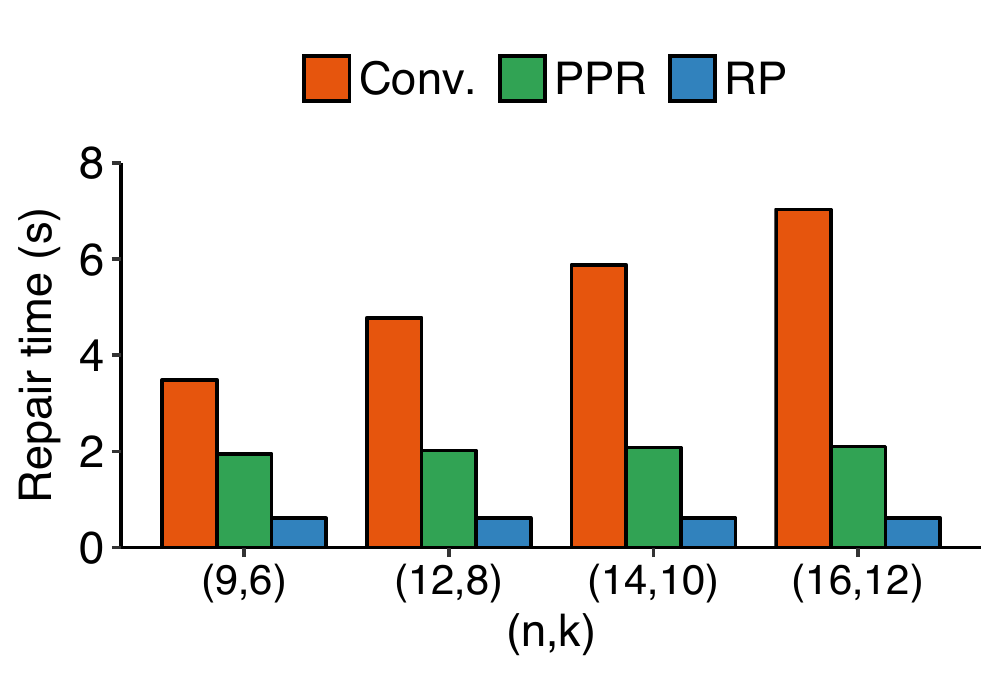} \\
\mbox{\small (a) Slice size} & 
\mbox{\small (b) Block size} &
\mbox{\small (c) Coding parameters} \\
\includegraphics[width=2.2in]{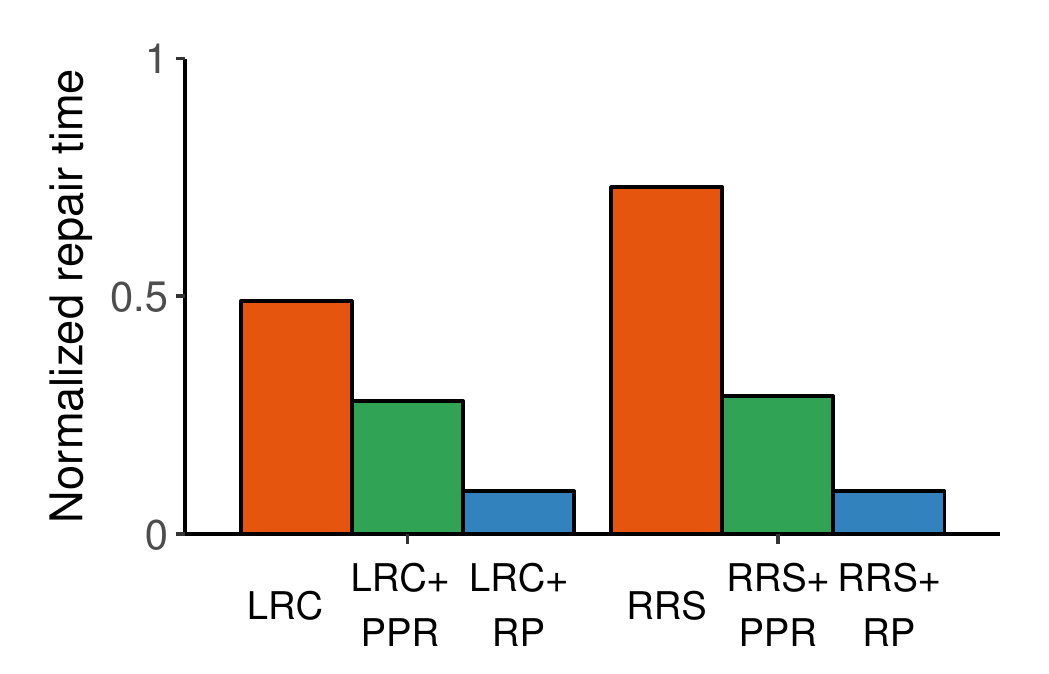} &
\includegraphics[width=2.2in]{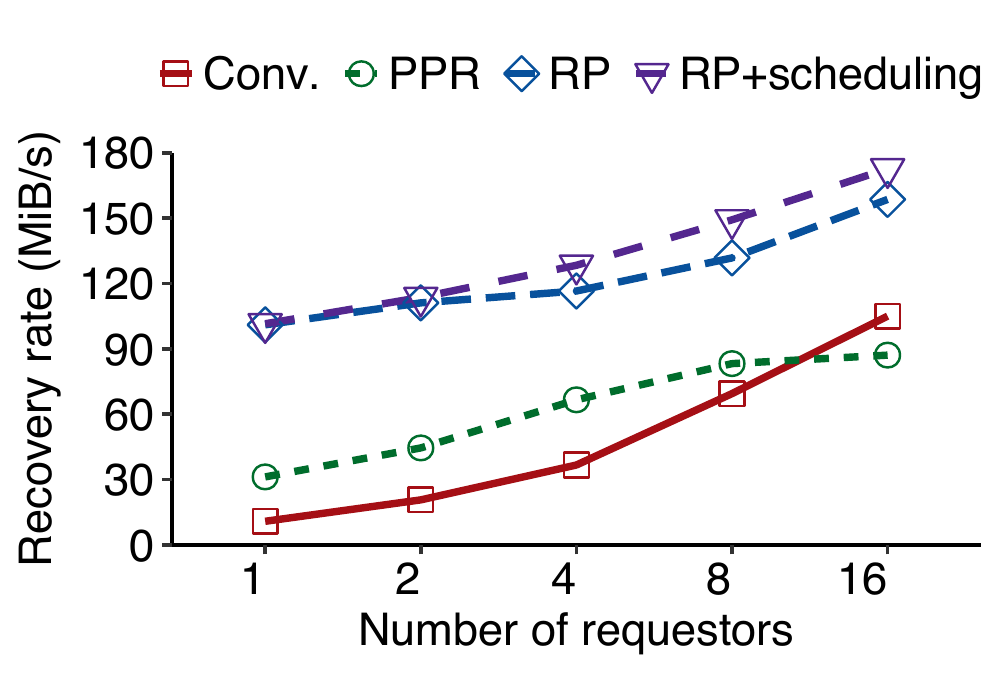} &
\includegraphics[width=2.2in]{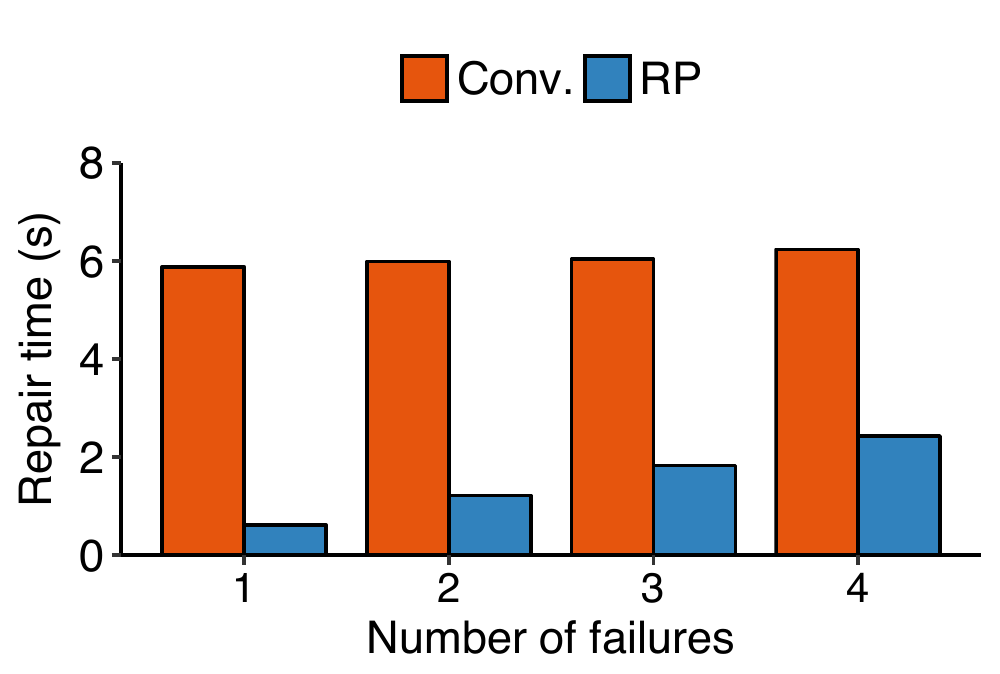} \\
\mbox{\small (d) Repair-friendly codes} &
\mbox{\small (e) Full-node recovery} &
\mbox{\small (f) Multi-block repair} \\
\includegraphics[width=2.2in]{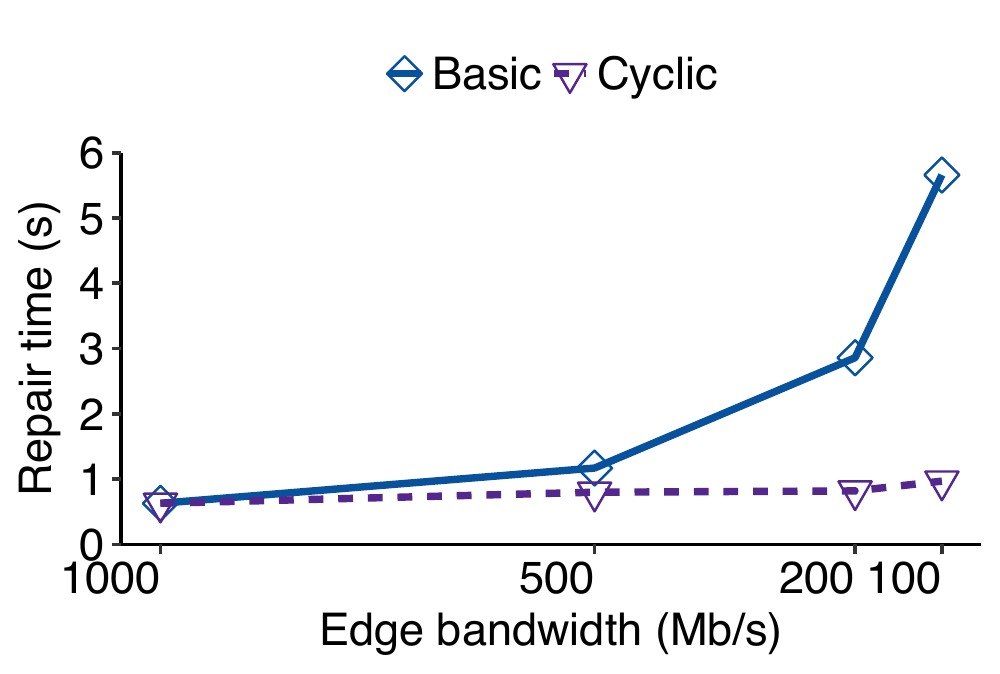} &
\includegraphics[width=2.2in]{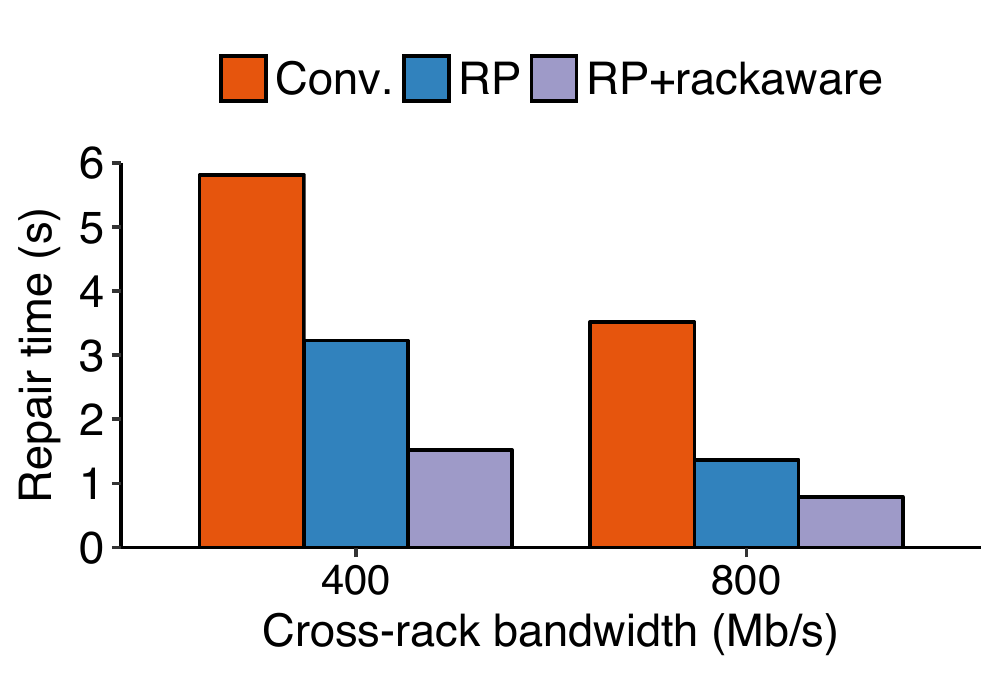} &
\includegraphics[width=2.2in]{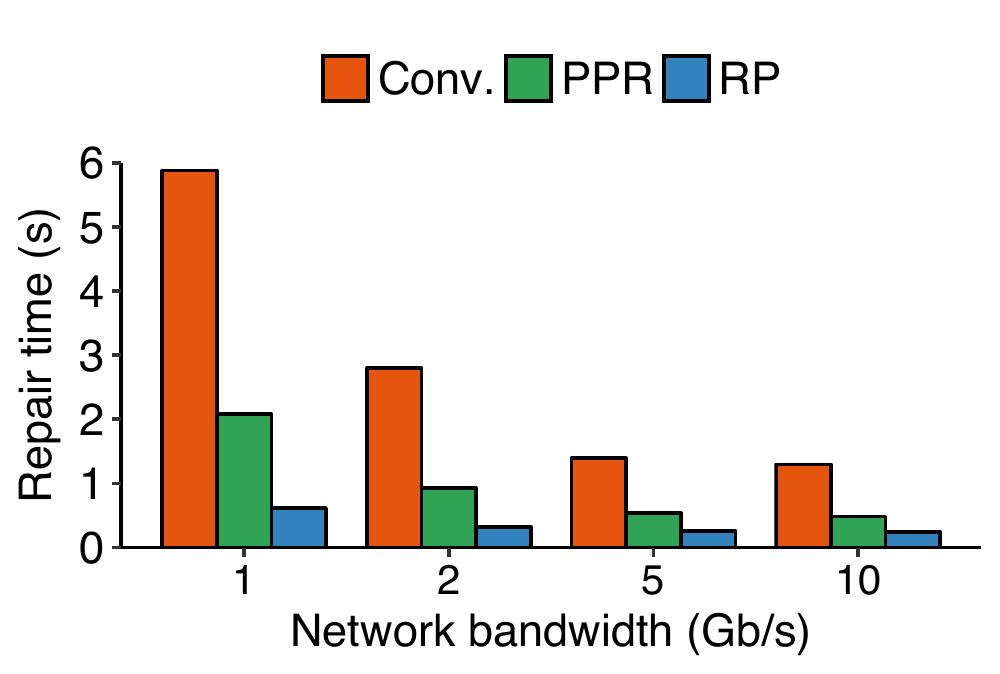} \\
\mbox{\small (g) Limited edge bandwidth} &
\mbox{\small (h) Rack awareness} &
\mbox{\small (i) Varying network bandwidth}
\end{tabular}
\caption{Evaluation on a local cluster.}
\label{fig:bench}
\end{figure*}

\noindent
{\bf Methodology:}
We first evaluate \sysname as a standalone system on a local cluster.  Our
local cluster comprises 17 machines, each of which has a quad-core 3.4\,GHz
Intel Core i5-3570 CPU, 16\,GiB RAM, and a Seagate ST1000DM003-1CH162 1\,TiB
SATA hard disk\footnote{Each machine in our local cluster has a faster CPU and
more RAM than the one used in our conference paper \cite{li17}.  Thus, we have
re-run all experiments and the values presented in this paper are different
from those in \cite{li17}.  Nevertheless, we still observe the significant
performance gain of repair pipelining.}.
We host the coordinator on one machine and 16 helpers on the remaining ones. 
By default, all machines are connected via a 1\,Gb/s Ethernet switch.  The
1\,Gb/s bandwidth can be viewed as modeling the cross-rack bandwidth available
for repair tasks in a production cluster \cite{sathiamoorthy13}, in which the
blocks of a stripe are stored in distinct racks.  We also connect the machines
via a 10\,Gb/s Ethernet switch and evaluate \sysname in higher network speeds
(Figures~\ref{fig:bench}(h) and \ref{fig:bench}(i)).

Initially, we store coded blocks in the local file system of each machine,
and load block locations and stripe information into the coordinator.  We
simulate a ``failed'' machine by erasing its stored blocks, and repair the
failed block of each stripe in a requestor.  We host the requestor on a
machine that does not store any available block of the repaired stripe, so as
to ensure that the available blocks are always transmitted over the network.
By default, we configure 64\,MiB block size, 32\,KiB slice size, and (14,10)
RS codes; note that (14,10) RS codes are also used by Facebook
\cite{sathiamoorthy13,rashmi14}.  We vary one of the settings at a time and
evaluate its impact.

We mainly compare the basic version of repair pipelining described in
\S\ref{sec:design} with conventional repair (\S\ref{sec:background}) and PPR
\cite{mitra16}.  We focus on three key repair metrics:
\begin{itemize}
\item
{\bf Single-block repair time:} the latency from issuing a degraded read
request to a failed block until the block is reconstructed; 
\item
{\bf Full-node recovery rate:} the ratio of the amount of recovered data in a
failed node to the total repair time; and
\item
{\bf Multi-block repair time:} the latency from issuing a request of repairing
multiple failed blocks in a stripe until they are all reconstructed.  
\end{itemize}
All results are averaged over 10 runs.  We find that the standard deviations
are small and hence omit them from the plots.

\paragraph{Slice size:}
Figure~\ref{fig:bench}(a) shows the single-block repair time versus the slice
size in repair pipelining; for fair comparisons, we also partition the blocks
into 32\,KiB slices in both conventional repair and PPR, so that they can also
exploit parallelism for better performance.  We further plot the transmission
time of directly sending a single block over a 1\,Gb/s link (labeled as
``Direct send'').  From the figure, we see that repair pipelining shows high
repair time when the slice size is small, even though more slices are
pipelined during a repair (i.e., $s$ is small).  The reason is that the
overhead of issuing transmission requests for many slices becomes significant.
We see that the repair time decreases as the slice size increases up to
32\,KiB (where $s=$~2,048), and then increases since there are too few slices
in a block being pipelined (i.e., less parallelization).  When the slice size
is 32\,KiB, repair pipelining reduces the single-block repair time by 89.5\%
and 69.5\% compared to conventional repair and PPR, respectively.  

Also, the direct send time of transferring a 64\,MiB block is 0.57s, which is
almost network-bound in our 1\,Gb/s network. The single-block repair time of
repair pipelining is only 8.8\% more than the direct send time. This shows the
feasibility of reducing the single-block repair time to almost the same as the
normal read time for a single available block. 

\paragraph{Block size:} Figure~\ref{fig:bench}(b) shows the single-block
repair time versus the block size.  Repair pipelining reduces the single-block
repair time by 88.8-91.6\% and 66.0-91.8\% compared to conventional repair and
PPR, respectively.

\paragraph{Coding parameters:} Figure~\ref{fig:bench}(c) shows the
single-block repair time versus $(n,k)$.  The single-block repair time of
conventional repair significantly increases with $k$, while that of PPR also 
increases with $k$ (albeit less significantly than in conventional repair).
On the other hand, the single-block repair time of repair pipelining remains
almost unchanged.  As $k$ increases from 6 to 12, the repair time reduction
of repair pipelining increases from 82.5\% to 91.2\% compared to conventional
repair, and from 68.6\% to 70.4\% compared to PPR.

\paragraph{Repair-friendly codes:}  We demonstrate how repair pipelining is
compatible with practical erasure codes.  We consider two state-of-the-art
repair-friendly codes: Local Reconstruction Codes (LRC) \cite{huang12} and
Rotated RS codes \cite{khan12}.
LRC partitions the data blocks into local groups and associates a local parity
block with each local group of data blocks.  It improves the performance of a
single-block repair, which can now be done within a local group, at the
expense of higher storage redundancy.  On the other hand, Rotated RS codes
arrange the layout of parity blocks to improve the performance of a degraded
read to a series of data blocks.  We configure LRC with $k=$~12 data blocks
that are partitioned in two local groups with six blocks each, and Rotated RS
codes with $(n,k)=$~(16,12).  LRC reads only six available blocks within a
local group for repairing a failed block, while Rotated RS codes on average
read nine blocks for repairing a failed block.  

Figure~\ref{fig:bench}(d) shows the normalized single-block repair time
with respect to conventional repair of (16,12) RS codes.  The normalized
single-block repair time of repair pipelining (around 0.1) is much smaller
than those of LRC and Rotated RS codes by effectively utilizing the bandwidth
resources of all helpers.  We observe the same improvement in PPR, but its
repair time reduction is less than that of repair pipelining.

\paragraph{Full-node recovery:}  We now evaluate full-node recovery with
multiple requestors and our greedy scheduling in helper selection
(\S\ref{subsec:recovery}).  Specifically, we randomly write multiple stripes
of blocks across all 16 helpers in the local cluster.  We erase 64 blocks from
64 stripes (one block per stripe) in one helper to mimic a single node
failure, and recover all the erased blocks simultaneously.  We distribute the
reconstructed blocks evenly across a number of requestors, varied from one to
16. Each requestor is deployed in a distinct machine. 

We consider two cases of helper selection in repair pipelining: (i) we index
the helpers from 1 to 16, and always select the available blocks from the $k$
helpers that have the smallest indexes in a stripe for repair (labeled as
``RP''); and (ii) we use the greedy approach to select $k$ helpers that are
least recently accessed for repair (labeled as ``RP+scheduling'').  We also
evaluate conventional repair and PPR, both of which select helpers without
greedy scheduling.

Figure~\ref{fig:bench}(e) shows the full-node recovery rates.  As the number
of requestors increases, the recovery rates of all schemes increase.
Conventional repair sees the largest gain by distributing the repair load
across more requestors.  Interestingly, as the number of requestors increases
to 16, conventional repair even achieves a slightly higher recovery rate than
PPR.  However, repair pipelining still outperforms conventional repair by
making bandwidth utilization more balanced.  Furthermore, our greedy
scheduling achieves an observable gain when there are a large number of
requestors.  For example, when there are eight (resp. 16) requestors, the
recovery rate of repair pipelining without greedy scheduling is 1.89$\times$
(resp.  1.51$\times$) that of conventional repair, and our greedy scheduling
further increases the recovery rate of repair pipelining by 13.3\% (resp.
8.9\%).  

\paragraph{Multi-block repair:} Figure~\ref{fig:bench}(f) shows the
multi-block repair time versus the number of failed blocks in a stripe. Here, 
we compare repair pipelining and conventional repair only, and omit PPR as its
design does not address the multi-block repair of a stripe.  Conventional repair
has relatively stable repair time (ranging from 5.88\,s to 6.23\,s) regardless
of the number of failed blocks being repaired, as it always retrieves $k$
available blocks for repairing the failed blocks of a stripe.  On the other
hand, the repair time of repair pipelining almost increases linearly with the
number of failed blocks.  Nevertheless, repair pipelining still has 60.9\%
less repair time than conventional repair for a four-block repair.

\paragraph{Limited edge bandwidth:}  Our previous tests focus on homogeneous
environments, and we now move our evaluation to heterogeneous environments. 
We show the benefits of the cyclic version when a requestor sits at the
network edge and the edge bandwidth from the storage system to the requestor
is limited (\S\ref{subsec:parallel}).  Specifically, we use the Linux
\texttt{tc} command \cite{tc} to limit the edge bandwidth from each helper to
the requestor.  We compare the cyclic version with the basic version in
\S\ref{sec:design}. 

Figure~\ref{fig:bench}(g) shows the single-block repair time versus the edge
bandwidth.  When the edge bandwidth is 1\,Gb/s (i.e., the homogeneous case),
both the basic and cyclic versions have almost identical repair time.  As the
edge bandwidth decreases, the repair time of the basic version increases
significantly, while that of the cyclic version only increases mildly by
allowing the requestors to read the reconstructed data from multiple helpers
in parallel.  For example, the cyclic version has 82.8\% less repair time than
the basic version when the edge bandwidth is 100\,Mb/s.

\paragraph{Rack awareness:} We evaluate repair pipelining in a rack-based data
center scenario.  Specifically, we configure (9,6) RS codes.  We divide our
cluster into three logical racks, and use the Linux {\tt tc} command \cite{tc}
to limit the cross-rack bandwidth.  We distribute the $n=9$ blocks of each
stripe evenly across the three logical racks (i.e., $n/3=3$ blocks per rack),
so that the block placement can tolerate any single-rack failure. We
compare repair pipelining with and without rack awareness
(\S\ref{subsec:rackaware}), as well as conventional repair; we do not consider
PPR here as its design does not address rack awareness. 

Figure~\ref{fig:bench}(h) shows the single-block repair time in two
cross-rack bandwidth settings: 400\,Mb/s and 800\,Mb/s.  Repair pipelining
without rack awareness reduces the repair time of conventional repair, yet
with rack awareness, we observe a further drop of the single-block repair
time.  For example, when the cross-rack bandwidth is 800\,Mb/s, repair
pipelining without rack awareness reduces the single-block repair time of
conventional repair by 60.9\%; with rack awareness, the reduction further
improves to 77.6\% since the cross-rack repair traffic is minimized.

\paragraph{Varying network bandwidth:}  We evaluate repair pipelining when the
network bandwidth is above 1\,Gb/s, in which the computation and disk I/O
overheads become significant.  We now connect all machines in our local
cluster via a 10\,Gb/s Ethernet switch.  We use the Linux {\tt tc} command
\cite{tc} to vary the available network bandwidth of each node (up to
10\,Gb/s).  

Figure~\ref{fig:bench}(i) shows that single-block repair time
versus the network bandwidth.  As the available network bandwidth
increases, the single-block repair time decreases in all schemes.  Also, the
repair time reduction of repair pipelining also drops due to the more
significant overheads in both computation and disk I/O.  Nevertheless, repair
pipelining still shows a performance gain.  For example, when the network
bandwidth is 10\,Gb/s, repair pipelining reduces the single-block repair time
by 81.4\% and 50.0\% compared to conventional repair and PPR, respectively
(while the reduction reaches around 90\% and 70\% when the network bandwidth
is 1\,Gb/s, as shown in Figure~\ref{fig:bench}(a)). 

\subsection{Evaluation on Amazon EC2}
\label{subsec:eval_ec2}

\noindent
{\bf Methodology:}  We evaluate \sysname on Amazon EC2.  Specifically, we
consider geo-distributed clusters that span multiple geographic regions
\cite{ford10,aguilera13,chen17}, in which erasure-coded blocks are striped
across regions to protect against large-scale correlated failures.  We
evaluate \sysname on two Amazon EC2 clusters, one in North America and one in
Asia.  Table~\ref{tbl:heteroThpt} shows one of our \texttt{iperf} \cite{iperf}
measurement tests for the inner-region and cross-region bandwidth values on
Amazon EC2 across four regions respectively in North America and Asia.  We
observe that the inner-region bandwidth is in general more abundant than
the cross-region bandwidth, and the cross-region bandwidth has a high degree
of variance.

\begin{table}[t]
\centering
\begin{subtable}{.51\linewidth}
\centering
\small
\hspace{-15pt}
\begin{tabular}{c|cccc}
\hline
Bandwidth   & California   & Canada & Ohio   & Oregon    \\
\hline
California  & 501.3       & 57.2  & 44.1   & 299.9    \\
Canada      & 55.3        & 732.0 & 63.3   & 48.0     \\
Ohio        & 46.3        & 65.7  & 332.5  & 95.6     \\
Oregon      & 297.8       & 50.2  & 93.6   & 250.1    \\
\hline
\end{tabular}
\caption{North America}
\end{subtable}%
\begin{subtable}{.51\linewidth}
\centering
\small
\begin{tabular}{c|cccc}
\hline
Bandwidth   & Mumbai    & Seoul     & Singapore     & Tokyo \\
\hline
Mumbai      & 624.8     & 62.3     & 39.5         & 37.7  \\
Seoul       & 63.8      & 265.7    & 86.1         & 183.2 \\
Singapore   & 41.5      & 88.1     & 493.0        & 49.1  \\
Tokyo       & 39.7      & 181.0    & 46.9         & 489.1 \\
\hline
\end{tabular}
\caption{Asia}
\end{subtable}
\caption{An \texttt{iperf} test of inner- and cross-region bandwidth
measurements (in Mb/s) on Amazon EC2 in North America and Asia.  Each value is
the measured bandwidth from the row region to the column region. Note that
the bandwidth values fluctuate across different tests.}
\label{tbl:heteroThpt}
\end{table}

We deploy four EC2 instances per region per cluster to host helpers (i.e., 16
helpers in total), and one EC2 instance in Ohio and Singapore to host the
coordinator for the North America and Asia clusters, respectively.  Note that
the overhead of accessing the coordinator has negligible impact on the overall
repair performance.  We focus on evaluating the single-block repair time of a
degraded read issued by a requestor.  We host the requestor on an EC2 instance
in each region and study how the performance varies across regions.  All EC2
instances are of type \texttt{t2.micro}.

We configure 64\,MiB block size and 32\,KiB slice size for repair pipelining.
We use (16,12) RS codes and distribute the 16 blocks of each stripe across the
16 EC2 instances in four regions; this also provides fault tolerance against
any single-region failure.  We consider two versions of repair pipelining: the
basic version in \S\ref{sec:design} (labeled as ``RP''), which finds a random
path across $k$ randomly selected helpers, and the optimal version in
\S\ref{subsec:weighted} (labeled as ``RP+optimal''), which finds an optimal
path via Algorithm~\ref{alg:hetero}.  Note that the network bandwidth
fluctuates over time, although inner-region bandwidth remains higher than
cross-region bandwidth, as shown in Table~\ref{tbl:heteroThpt}.  Thus, the
optimal version probes the network bandwidth via {\tt iperf} before each run
of experiments.  We average our results over 10 runs, and also include the
standard deviations as the results have higher variances than in our local
cluster.

\begin{figure}[t]
\centering
\begin{tabular}{cc}
\includegraphics[width=2.9in]{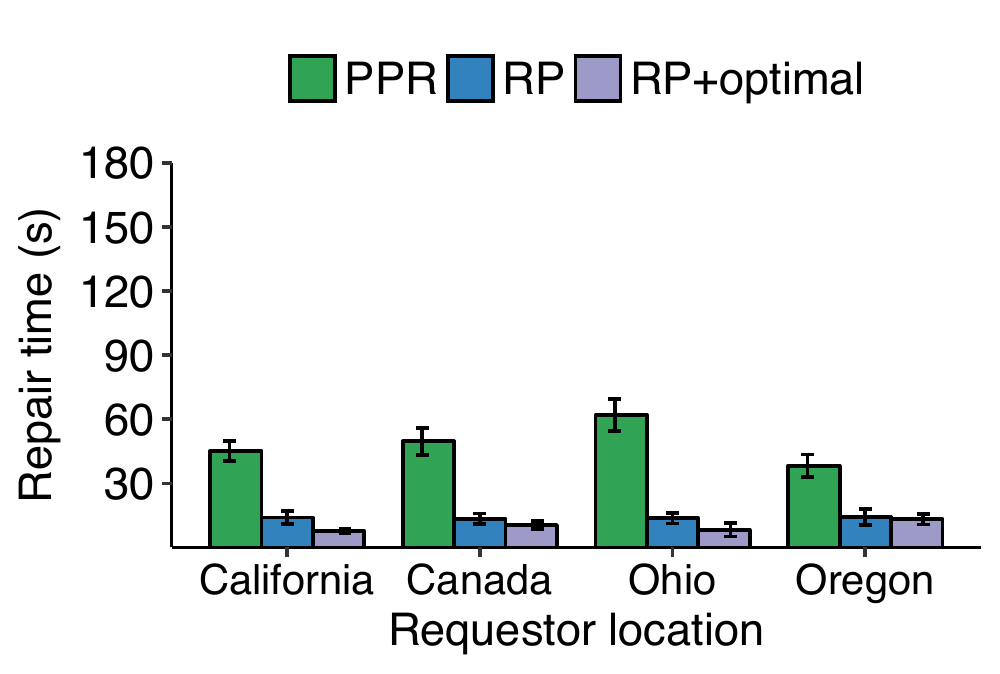} &
\includegraphics[width=2.9in]{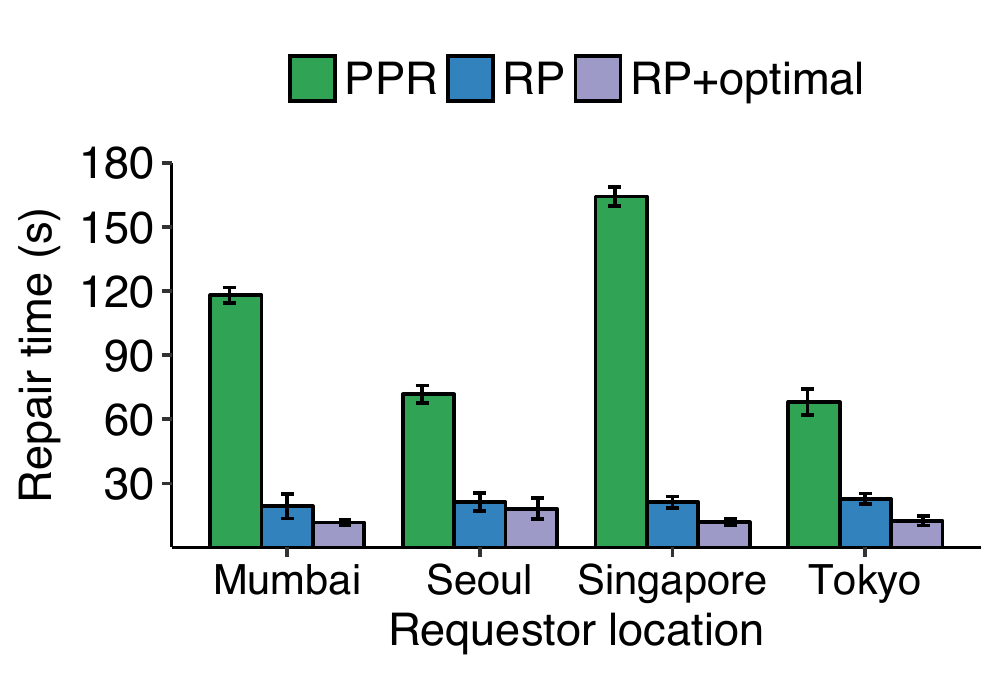}\\
\mbox{\small (a) North America cluster}&
\mbox{\small (b) Asia cluster}
\end{tabular}
\caption{Evaluation on Amazon EC2.}
\label{fig:expHetero}
\end{figure}

\paragraph{Results:}
Figure~\ref{fig:expHetero} shows the single-block repair time and the
standard deviations of PPR and the two versions of repair pipelining in both
clusters; we do not show the results of conventional repair, whose repair time
goes beyond 200\,s.  Repair pipelining (without weighted path selection)
achieves repair time saving over PPR in all cases when the requestor is in
different regions.  The repair time reduction is 62.7-78.0\% for North America
and 66.6-87.1\% for Asia.  Our weighted path selection further reduces the
repair time by 7.3-45.4\% for North America and 14.5-45.0\% for Asia, compared
to repair pipelining without weighted path selection. Note that our weighted
path selection can be done in around 1\,ms (\S\ref{subsec:weighted}), which is
negligible compared to the repair time in our evaluation.

\subsection{Evaluation on HDFS-RAID, HDFS-3, and QFS}
\label{subsec:eval_dfs}

\noindent
{\bf Methodology:} We evaluate the integration of \sysname into HDFS-RAID,
HDFS-3, and QFS, all of which are deployed in our local cluster
(\S\ref{subsec:eval_local}).  We co-locate a helper daemon with each of the 16
storage nodes (i.e., DataNodes in HDFS-RAID and HDFS-3, or ChunkServers in
QFS).  By default, we set the slice size of repair pipelining as 32\,KiB and
block size as 64\,MiB.  For HDFS-RAID and HDFS-3, we vary $(n,k)$, while for
QFS, we use its default (9,6) RS codes and vary the slice size and block size.
We consider three repair schemes: (i) the original repair implementations of
HDFS-RAID, HDFS-3 and QFS, all of which are based on conventional repair, (ii)
the conventional repair under \sysname, and (iii) the basic version of repair
pipelining in \S\ref{sec:design} under \sysname.

For HDFS-RAID and QFS, we evaluate degraded reads (in single-block repair
time) issued by a requestor that is attached with either an HDFS-RAID client
or a QFS ChunkServer.  For HDFS-3, 
we observe similar results of the single-block repair time as in HDFS-RAID.
Thus, we focus on evaluating full-node recovery in HDFS-3,
in which we evenly distribute 64 stripes of blocks across all DataNodes,
followed by erasing all blocks of a DataNode and repairing the lost blocks in
a new DataNode.  We report the averaged results over 10 runs as in
\S\ref{subsec:eval_local} (the standard deviations are small and omitted).

\begin{figure*}[t]
\centering
\begin{tabular}{cc}
\includegraphics[width=2.9in]{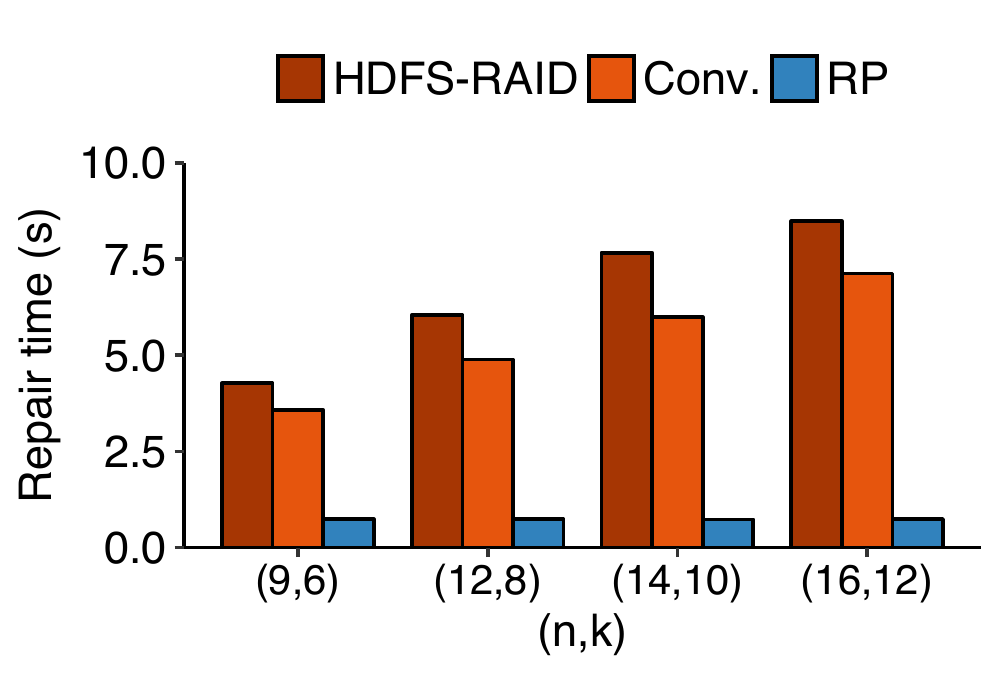}&
\includegraphics[width=2.9in]{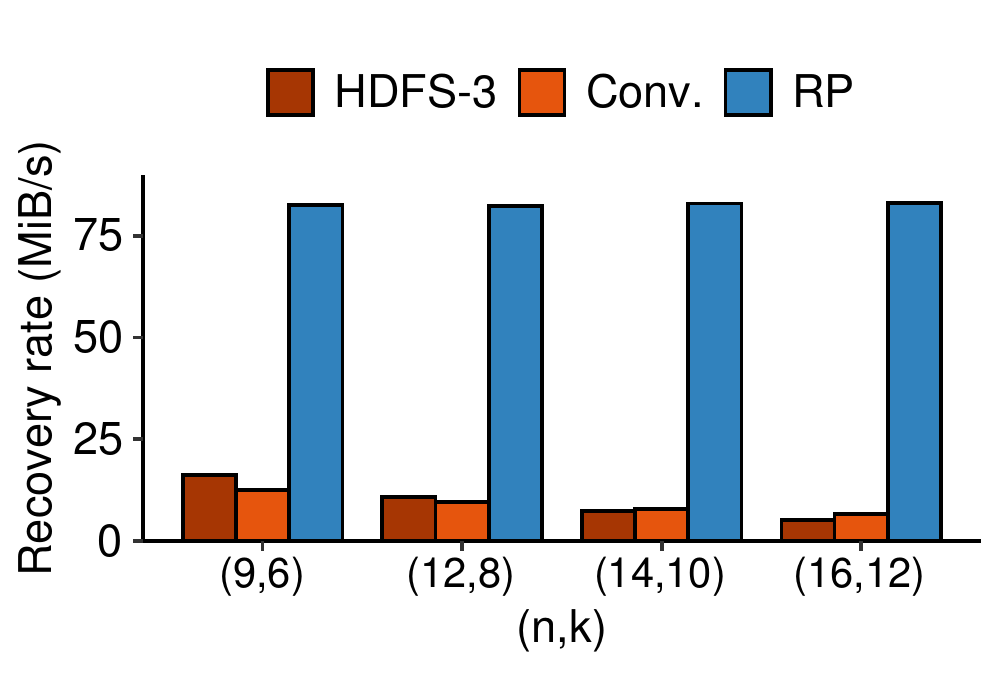}\\
\parbox[t]{2.4in}{\small(a) HDFS-RAID: single-block repair time versus coding
parameters} &
\parbox[t]{2.4in}{\small(b) HDFS-3: full-node recovery rate versus coding parameters}
\vspace{3pt} \\
\includegraphics[width=2.9in]{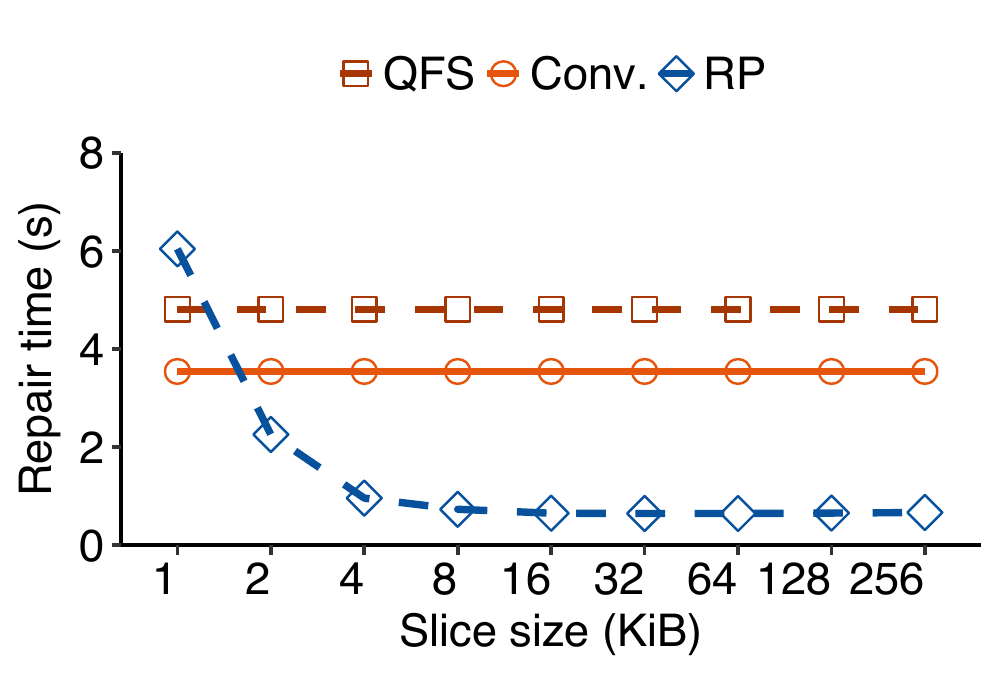}&
\includegraphics[width=2.9in]{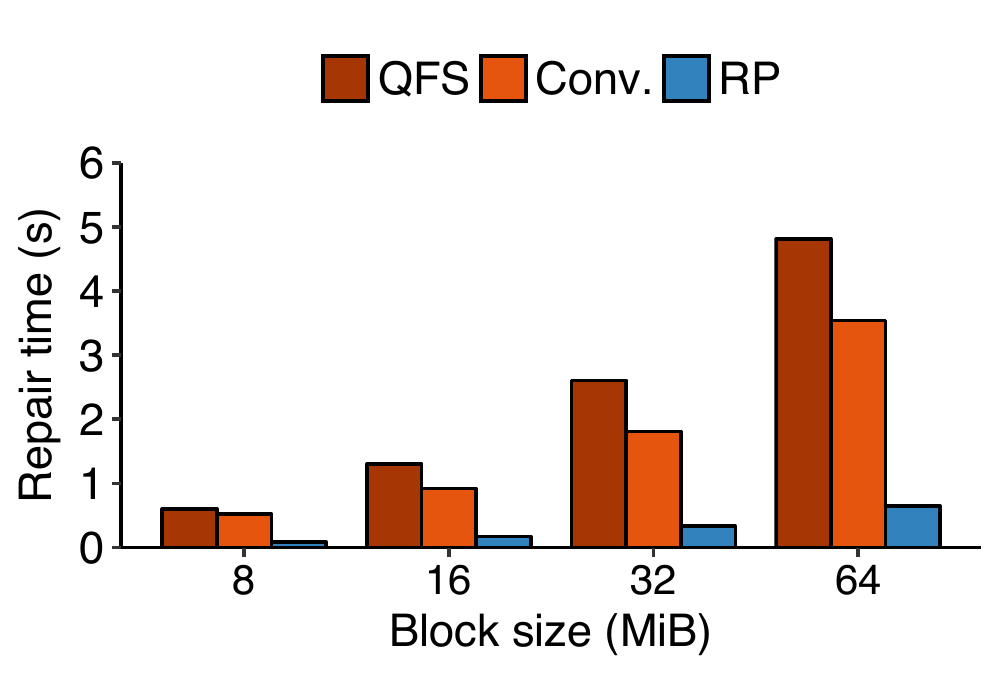}\\
\parbox[t]{2.4in}{\small(c) QFS: single-block repair time versus slice size}&
\parbox[t]{2.4in}{\small(d) QFS: single-block repair time versus block size}
\end{tabular}
\caption{Evaluation on HDFS-RAID, HDFS-3, and QFS.}
\label{fig:exp}
\end{figure*}

\paragraph{Results:}
Figure~\ref{fig:exp} shows the evaluation results.  First, repair pipelining
under \sysname significantly improves the repair performance of the original
repair implementations of HDFS-RAID, HDFS-3, and QFS.  Specifically, for
HDFS-RAID, repair pipelining reduces the single-block repair time by
82.7-91.2\% for different $(n,k)$ (Figure~\ref{fig:exp}(a)); for HDFS-3, it
achieves 5.1-16.0$\times$ full-node recovery rate for different $(n,k)$
(Figure~\ref{fig:exp}(b)); for QFS, it reduces the single-block repair time by 
up to 86.6\% when the slice size is 32\,KiB and the block size is 64\,MiB
(Figures~\ref{fig:exp}(c) and \ref{fig:exp}(d)). 

We observe that moving the repair logic to \sysname improves single-block
repair performance.  Specifically, conventional repair under \sysname reduces
the single-block repair time by up to 21.8\% and 26.3\% in HDFS-RAID and QFS,
respectively, compared to the original conventional repair implementation. The
reason of the performance gain is that the helpers of \sysname can directly
access the stored blocks via the native file system, instead of fetching the
blocks through the distributed storage system routine.  For full-node
recovery, conventional repair under \sysname outperforms the original
conventional repair when $k$ is large (i.e., $k=$~10 or 14).  The reason is
that when $k$ increases, the overhead of initiating connections to $k$
DataNodes for retrieving available blocks in HDFS-3 also increases.
Nevertheless, we emphasize that the repair performance gain mainly comes from
repair pipelining, rather than the implementation of \sysname.  Although
moving repair to \sysname reduces the repair time, the reduction is minor
compared to the reduction achieved by repair pipelining.

\subsection{Evaluation of Different Repair Pipelining Implementations}
\label{subsec:eval_rp}

\noindent
{\bf Methodology:}
We compare different repair pipelining implementations based on \sysname in
our local cluster (\S\ref{subsec:eval_local}). 
First, we compare the block-level and slice-level repair pipelining
approaches, and implement two baseline variants called {\em Pipe-B} and 
{\em Pipe-S}. Pipe-B implements block-level repair pipelining (i.e., without
slicing) along a linear path of helpers, each of which sends a partially
repaired block to the next helper (or the requestor for the last helper); in
essence, Pipe-B is the na\"ive approach as described in
\S\ref{subsec:degraded}. Pipe-S implements slice-level repair pipelining
without parallelization.  It realizes the sub-operations of repairing a slice
inside a helper (i.e., receiving the partially repaired slice from the
preceding helper, reading its locally stored slice, computing a new partially
repaired slice, and sending the partially repaired slice) in a {\em serial}
manner. 
Our current repair pipelining implementation (referred to as {\em RP}) also
performs slice-level repair pipelining.  Compared to Pipe-S, RP carefully
parallelizes the sub-operations of slices in each helper to simultaneously
utilize all available resources.  Note that RP is our default implementation
used in the previous experiments 
(\S\ref{subsec:eval_local}-\S\ref{subsec:eval_dfs}). 

We also compare our repair pipelining implementation with the design in PUSH
\cite{huang15} in full-node recovery as described in \S\ref{subsec:eval_local}
(note that PUSH does not consider single-block repair).  As the source code of
PUSH is unavailable, we implement the two variants of PUSH, namely PUSH-Rep
and PUSH-Sur, in \sysname based on the description of the paper
\cite{huang15}, which we refer to them as {\em Pipe-Rep} and {\em Pipe-Sur},
respectively.  Both Pipe-Rep and Pipe-Sur implement block-level repair
pipelining.  Pipe-Rep reconstructs all failed blocks in a single node, while
Pipe-Sur distributes the reconstructed blocks across all 16 nodes in our local
cluster in a round-robin manner.  For comparisons, we consider two variants of
our repair pipelining implementation (with greedy scheduling enabled), namely
{\em RP-single} and {\em RP-all}. RP-single reconstructs all failed blocks in
a single node, while RP-all distributes all reconstructed blocks across all 16
nodes in our local cluster.  Both RP-single and RP-all are configured with 16
requestors: for RP-single, we deploy all 16 requestors in the node where the
failed blocks are reconstructed; for RP-all, we deploy one requestor per node.
Compared to PUSH-Rep and PUSH-Sur, both RP-single and RP-all implement
slice-level repair pipelining.  

\begin{figure*}[t]
\centering
\begin{tabular}{@{\ }c@{\ }c}
\includegraphics[width=2.9in]{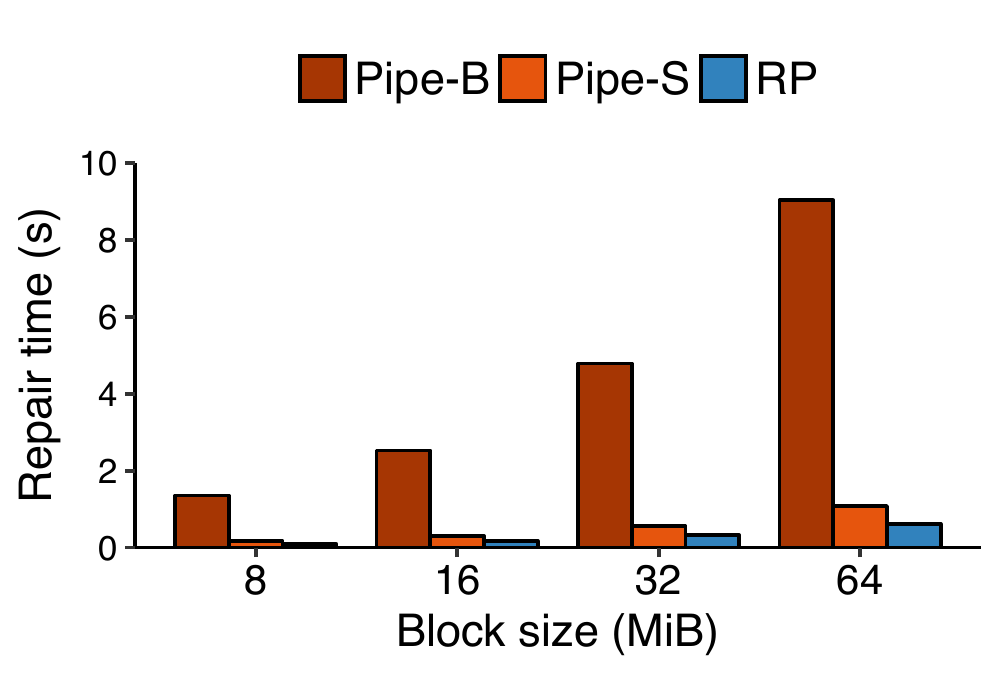} &
\includegraphics[width=2.9in]{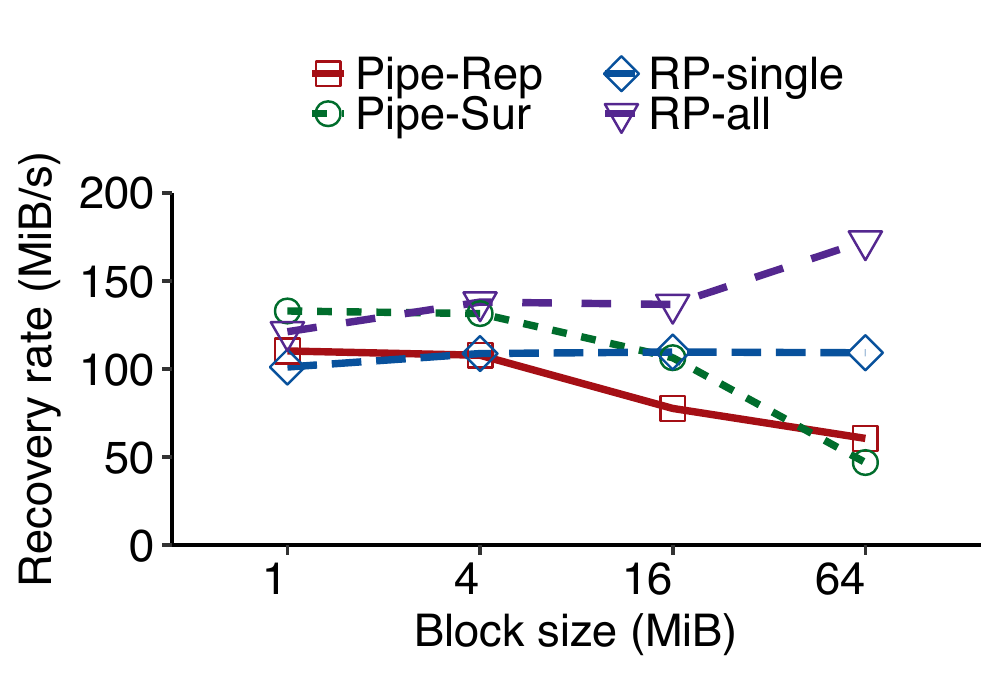}\\
\parbox[t]{2.4in}{\small (a) Single-block repair time versus block size} & 
\parbox[t]{2.4in}{\small (b) Full-node recovery rate versus block size}
\end{tabular}
\caption{Evaluation on different repair pipelining implementations.}
\label{fig:rpimpl}
\end{figure*}

\paragraph{Results:} Figure~\ref{fig:rpimpl} shows the results, averaged over
10 runs.  Figure~\ref{fig:rpimpl}(a) evaluates the single-block repair
time versus the block size for Pipe-B, Pipe-S, and RP, where both Pipe-S and
RP have the slice size fixed as 32\,KiB.  Pipe-B has the largest single-block
repair time (e.g., 9.0\,s for repairing a 64\,MiB block), while Pipe-S
significantly reduces the single-block repair time (e.g., 1.1\,s for repairing
a 64\,MiB block) through slice-level repair pipelining.  This again shows that
slice-level repair pipelining can improve the repair performance via more
fine-grained parallelization.  RP further reduces the single-block repair time
(e.g., 0.61\,s for repairing a 64\,MiB block) by carefully scheduling the
slice-level repair sub-operations in a parallel fashion.  Overall, RP reduces
the single-block repair time of Pipe-S by 41.1-43.0\% across all block sizes.
Note that we observe similar performance differences between Pipe-S and RP for
different slice sizes, so we omit the results here. 

Figure~\ref{fig:rpimpl}(b) evaluates the full-node recovery rate versus the
block size for Pipe-Rep, Pipe-Sur, RP-single, and RP-all. Here, we repair
4\,TiB of lost data (i.e., the number of reconstructed blocks is 4\,TiB
divided by the block size), and fix the slice size for RP-single and RP-all as 
32\,KiB.  We observe that when the block size is 1\,MiB, both Pipe-Rep and
Pipe-Sur have higher recovery rates than RP-single and RP-all by 9.0\% and
9.7\%, respectively.  The reason is that Pipe-Rep and Pipe-Sur benefit from
block-level repair pipelining across a large number of blocks in small block
sizes.  On the other hand, in RP-single and RP-all, each block is only divided
into a limited number of slices (e.g., 32 slices for a 1\,MiB block).  They do
not benefit much from slice-level repair pipelining. 

Nevertheless, as the block size increases, the recovery rates of both Pipe-Rep
and Pipe-Sur drop significantly, as the number of blocks decreases for larger
block sizes and the performance gain from pipelining is limited.  On the other
hand, the recovery rates of RP-single and RP-all increase with the block size,
as each block can now be divided into more slices.  Both RP-single and RP-all
can benefit from the slice-level repair pipelining for each block; note that 
they also allow multiple requestors to reconstruct the lost blocks in
parallel.  When the block size is 64\,MiB, the recovery rates of RP-single and
RP-all are 80.2\% and 268.1\% higher than those of Pipe-Rep and Pipe-Sur,
respectively.  Also, RP-all has a higher recovery rate than RP-single (e.g.,
by 58.1\% when the block size is 64\,MiB) by distributing the repair load
across the requestors in different nodes.   In summary, our current repair
pipelining implementation maintains its high performance gain in large block
sizes, which are commonly found in state-of-the-art distributed storage
systems (e.g., 64\,MiB \cite{ghemawat03} or 256\,MiB \cite{rashmi14}). 

\section{Related Work}
\label{sec:related}

Many new erasure codes have been proposed in the literature 
to mitigate repair overhead,
especially for a single-node repair.  To name a few, regenerating codes
\cite{dimakis10} minimize the repair traffic by allowing storage nodes to send
encoded data during a single-node repair.  Rotated RS codes \cite{khan12}
reduce the repair traffic and disk I/O of a degraded read to a sequence of
data blocks.  Hitchhiker \cite{rashmi14} extends RS codes \cite{reed60} to
piggyback parity information of one stripe into another stripe, and is shown
to reduce the repair traffic and I/O by up to 45\%.  PM-RBT codes
\cite{rashmi15} are special regenerating codes that simultaneously minimize
the repair traffic, disk I/O, and storage redundancy.  Butterfly codes
\cite{pamies16} are systematic regenerating codes that provide double-fault
tolerance.  Clay codes \cite{vajha18} couple multiple layers of MDS codes and
achieve optimality in terms of the repair traffic, disk I/O, storage
redundancy, as well as the sub-packetization level (i.e., the number of
sub-blocks divided within a block).  Locally repairable codes
\cite{huang12,sathiamoorthy13} add local parity blocks to mitigate repair I/O
with extra storage redundancy.  

Instead of constructing new erasure codes, we design new repair strategies for
general practical erasure codes.  Some prior studies are also along this
direction.   Tree-structured data regeneration \cite{li10} specifically
targets regenerating codes \cite{dimakis10}, and constructs a spanning tree
that maximizes the bandwidth utilization during repair.  
Lazy repair \cite{bhagwan04,silberstein14} defers immediate repair action
until a tolerable limit is reached.   To speed up full-node recovery, the
repair of multiple stripes can be parallelized across available nodes, as also
adopted by replicated storage \cite{chun06,ongaro11} and de-clustered RAID
arrays \cite{holland92}.  Degraded-first scheduling \cite{li14} targets
MapReduce on erasure-coded storage by scheduling map tasks to fully utilize
bandwidth in degraded reads.  FastPR \cite{shen19} reconstructs
in advance the data stored in soon-to-fail nodes, so as to speed up the repair
operation.
A closely related work to ours is PPR \cite{mitra16}, which reduces the
single-block repair time from $k$ timeslots to $\lceil\log_2(k+1)\rceil$
timeslots by parallelizing partial repair operations across different nodes,
while repair pipelining reduces the single-block repair time to one timeslot.

Several studies improve the repair performance of erasure-coded storage for
hierarchical data centers. 	Some studies \cite{hu17,prakash18,hou19} propose
new regenerating codes that minimize the cross-rack repair traffic for
hierarchical topologies, in which storage nodes are organized in racks.
CAR \cite{shen16} minimizes the cross-rack repair traffic for RS-coded storage
by first computing partial repaired results in each rack and then sending the
partial repaired results across racks. LAR \cite{xu19}, similar to CAR
\cite{shen16}, also studies how to minimize the cross-rack repair traffic in
the network core of a hierarchical topology by solving for a minimum spanning
tree.  ClusterSR \cite{shen20} not only minimizes the cross-cluster repair
traffic in geo-distributed storage, but also balances the upload and download
traffic in full-node recovery.
With hierarchy awareness, repair pipelining preserves the minimum cross-rack
repair traffic and further reduces the single-block repair time
(\S\ref{subsec:rackaware}).

Some studies also propose pipelined approaches to improve the repair
performance in erasure-coded storage.  PUSH \cite{huang15} forms a
reconstruction chain along different helpers and performs block-level repair
pipelining for full-node recovery.  In contrast, our repair pipelining design
schedules a single-block repair at a more fine-grained slice level, and we
show that how it substantially reduces the single-block repair time to almost
the same as the normal read time for a single block.  
Compared to PUSH, our contributions include: (i) by slicing a block (a
read/write unit of a distributed storage system) into smaller units, we show
that repair pipelining can reduce the degraded read time for an unavailable
block to almost the same as the normal read time for an available block (in
contrast, PUSH only focuses on full-node recovery); (ii) we present extensions
of repair pipelining for heterogeneous environments; (iii) we show how repair
pipelining can be readily integrated via \sysname into existing distributed
storage systems (i.e., HDFS-RAID, HDFS-3, and QFS); and (iv) we compare
different repair pipelining implementations (including our own implementation
of PUSH \cite{huang15}). 
LAR \cite{xu19} implements pipelined reconstruction by dividing blocks into
packets (i.e., slices in our case).  However, it does not formally analyze how
the packets are scheduled to minimize single-block repair time.  Parallel
Pipeline Tree (PPT) \cite{bai19} constructs an optimized repair tree based on
repair pipelining for heterogeneous environments, while we address some
special cases of heterogeneous environments, such as the scenario where the
link bandwidth between the storage system and the requestor is limited as well
as hierarchical data centers.  Our recent work, OpenEC \cite{li19}, provides a
framework that simplifies the deployment of repair pipelining through a
directed-acyclic-graph abstraction. 

\section{Conclusions}
\label{sec:conclusions}

Repair pipelining is a general technique that reduces the single-block repair
time to almost the same as the normal read time for a single available block
in erasure-coded storage.  It schedules the repair of a failed block across
storage nodes in units of slices in a pipelined manner, so as to evenly
distribute the repair traffic and fully utilize bandwidth resources across
storage nodes.  Our contributions include: (i) the design of repair pipelining
for both degraded reads and full-node recovery, (ii) the extensions of repair
pipelining with parallel reads, hierarchy awareness, and weighted path
selection for heterogeneous environments, (iii) the extension of repair
pipelining for repairing multiple failed blocks in a stripe, (iv) a repair
prototype \sysname and its integrations into HDFS-RAID, HDFS-3, and QFS, and
(v) the local cluster and Amazon EC2 experiments that show the repair speedup
through repair pipelining. 

\bibliographystyle{abbrv}
\bibliography{reference}

\end{document}